\documentclass[epsfig,12pt,onecolumn]{article}
\usepackage{amsfonts}
\usepackage{amssymb}
\usepackage{amsmath}
\usepackage{multicol}
\usepackage{graphicx}
\usepackage{float}
\usepackage{caption}

\makeatletter \@addtoreset{equation}{section}
\renewcommand{\theequation}{\thesection.\arabic{equation}}
\def \be{\begin{equation}}
\def \ee{\end{equation}}
\def \bea{\begin{eqnarray}}
\def \eea{\end{eqnarray}}

\newcommand{\nc}{\newcommand}
\nc{\al}{\alpha} \nc{\bib}{\bibitem} \nc{\la}{\lambda}
\nc{\C}{\mbox{\hspace{1.24mm}\rule{0.2mm}{2.5mm}\hspace{-2.7mm} C}}
\nc{\R}{\mbox{\hspace{.04mm}\rule{0.2mm}{2.8mm}\hspace{-1.5mm} R}}
\input{tcilatex}
\begin{document}

\title{\textbf{On Exceptional 't Hooft Lines in 4D-Chern-Simons Theory}}
\author{Youssra Boujakhrout\thanks{%
youssra\_boujakhrout@um5.ac.ma} \ and El Hassan Saidi\thanks{%
e.saidi@um5r.ac.ma} \\
{\small 1. LPHE-MS, Science Faculty}, {\small Mohammed V University in
Rabat, Morocco}\\
{\small 2. Centre of Physics and Mathematics, CPM- Morocco}}
\maketitle

\begin{abstract}
We study 't Hooft lines and the associated $\mathcal{L}$- operators in
topological 4D Chern-Simons theory with gauge symmetry given by the
exceptional groups E$_{6}$ and E$_{7}$. We give their oscillator
realisations and propose topological gauge quivers encoding the properties
of these topological lines where Darboux coordinates are interpreted in
terms of topological fundamental matter. Other related aspects are also
described.\newline
\textbf{Keywords:} 4D Chern-Simons theory, Wilson and 't Hooft lines,
Topological quivers.
\end{abstract}


\section{Introduction}

The discovery of four dimensional Chern-Simons theory \textrm{\cite{1A}} has
given a great impulse towards deep understanding of quantum integrability in
2D field theory \textrm{\cite{2A}} and integrable spin models \textrm{\cite%
{1B,2B,3B}} in lower dimensions. While standard Chern-Simons (CS) gauge
theories are topological theories involving hermitian gauge fields in odd
spacetime \textrm{\cite{1C}}, the Costello-Witten Yamazaki (CWY) theory
lives in 4D space and goes beyond the hermiticity property. This non unitary
feature allowed to extend the application of methods of standard QFT to
complexified gauge fields living on complex manifolds \textrm{\cite%
{1A,2A,1D,2D,3D}}. The 4D- CS theory gives a new approach to describe major
elements and phenomena of 2D integrable systems in terms of a complex CS
gauge potential living on $\Sigma \times \mathbb{C}$ and valued in a complex
Lie algebra. Here, the $\Sigma $ is a real 2D topological surface and $%
\mathbb{C}$ the complex holomorphic line \textrm{\cite{1A}}. Roughly
speaking, the CWY theory permits to represent the worldline $\mathrm{\gamma }%
_{z}$ of a particle in $\Sigma $ by a Wilson line operator W$_{\mathrm{%
\gamma }_{z}}$ characterized by: $\left( i\right) $ a complex spectral
parameter $z\in \mathbb{C},$ interpreted in 2D as rapidity of the particle 
\textrm{\cite{1E}}; and\emph{\ }$\left( ii\right) $ a representation space $%
V $ of the complex gauge group\emph{\ }$G$ describing intrinsic degrees of
freedom of the particle. In this framework, the crossing of two Wilson lines
in 4D is expressed by the famous R-matrix $\mathfrak{R}(z):V\otimes
V\rightarrow V\otimes V$\ that verifies the Yang-Baxter equation of
integrability.\emph{\ }The 4D topological CWY approach was also shown to
describe other topological objects like the ones we are interested in this
study namely 't Hooft lines \textrm{\cite{1F}}. They correspond to the
Baxter Q-operator of integrable spin chains \cite{2G,2GA,2GB} and have an
interesting interpretation in 4D CS theory. Their coupling to W$_{\mathrm{%
\gamma }_{z}}$ is given by the so- called $\mathcal{L}$-operator \cite%
{1F,1G,3G} playing a quite similar role as the R-matrix \cite{1A,2A}. This
gauge invariant quantity was recently interpreted as an observable measuring
the parallel transport of the gauge field sourced by the 't Hooft line and
was realized for A-type and D- type gauge symmetries using \textrm{%
symplectic oscillators \cite{1F,1G}.}\newline
In this paper, we contribute to 4D- CS theory with minuscule 't Hooft lines
by working out the oscillator realisation of the $\mathcal{L}$- operator for
the exceptional gauge symmetries; thus completing results obtained in \cite%
{1F,1B,1H}. First, we revisit useful aspects on 't Hooft lines in the 4D- CS
theory for generic gauge symmetries G while illustrating the construction
for $G=SL\left( N\right) $. Then, we focuss on the topological E$_{6}$ and E$%
_{7}$ theories and give the missing oscillator representation of the $%
\mathcal{L}$- operators associated with these gauge symmetries. Our $%
\mathcal{L}_{E_{6}}$ and $\mathcal{L}_{E_{7}}$ can be also viewed as a
generalization of the results regarding A- type\textrm{\  \cite{1H}} and D-
type \textrm{\cite{1B}} gauge groups. After that, we borrow ideas from
supersymmetric quiver gauge theories to propose a quiver representation of
the 't Hooft line operators. In this view, the $\mathcal{L}_{E_{s}}$'s are
represented by topological gauge quivers Q$_{E_{s}}$ where nodes and links
are interpreted in terms of topological gauge matter. To fix ideas on the
shape of these quivers, see the Figures \textbf{\ref{so10}} and \textbf{\ref%
{2828}}. In these topological Q$_{E_{s}}$'s, the Darboux coordinates of the
phase space of the $\mathcal{L}_{E_{s}}$'s are interpreted in terms of
topological fundamental matter and nodes as self-dual matter.\newline
The organisation of this study is as follows. In section 2, we introduce the
minuscule 't Hooft lines in 4D CS theory and describe some of their
representations as well as their implementation in the CWY theory. In
sections 3 and 4, we study the 4D- CS theory with E$_{6}$ gauge symmetry.
First, we build the $\mathcal{L}_{E_{6}}$ operator using 16 oscillators.
Then, we give our proposal regarding the topological gauge quiver Q$_{E_{6}}$%
. In section 5, we do the same thing for the the 4D- CS with gauge symmetry E%
$_{7}$. Section 6 is devoted to conclusion and perspectives. The last
section is an appendix detailing technical steps in the calculation of the
Lax operator for the E$_{7}$ theory.

\section{'t Hooft lines in 4D Chern-Simons theory}

In this section, we revisit some basic ingredients regarding classical 't
Hooft lines $\mathrm{\gamma }_{{\small z}}$ in 4D- CS theory with a rank r
ADE gauge symmetry G. First, we introduce these magnetically charged lines
with some charge $\mu $ to be specified later. For convenience, we refer to
these lines like tH$_{\mathrm{\gamma }_{{\small z}}}^{\mu }$. Then, we
describe their realisation in 4D topological CS theory by following the
Costello- Gaiotto- Yagi representation given in \textrm{\cite{1F}}. We refer
to this realisation as the CGY $\mathcal{L}$-operator. We end this review by
describing the phase space of this gauge invariant observable formulated as
an "RTT" equation and its relationship with Darboux coordinates of
symplectic geometry.

\subsection{Minuscule 't Hooft lines and CGY observable}

Generally speaking, the 't Hooft lines are a dual version of Wilson lines
introduced to represent the worldline of an infinitely heavy magnetic
monopole in spacetime \textrm{\cite{1I,2I,3I}}. In the 4D CS theory on $%
\Sigma \times \mathbb{C}$ with gauge symmetry G, a 't Hooft line tH$_{%
\mathrm{\gamma }_{{\small z}}}^{\mu }$ is a 1D topological defect $\mathrm{%
\gamma }_{{\small z}}$\ that lives in the topological surface\ $\Sigma $,
taken below as the plane $\mathbb{R}^{2}$.\emph{\ }As for the electrically
charged Wilson lines W$_{\xi _{w}}^{q_{e}}$, the magnetically charged tH$_{%
\mathrm{\gamma }_{{\small z}}}^{q_{m}}$ is also characterised by a spectral
parameter $z$ interpreted in the 4D CS theory as the position of the line
defect in the holomorphic sector $\mathbb{C}.$ An interesting family of the
tH$_{\mathrm{\gamma }_{{\small z}}}^{q_{m}}$ lines is given by the so-called 
\emph{minuscule 't Hooft lines} tH$_{\mathrm{\gamma }_{{\small z}}}^{\mu }$\
on which we will be focussing on in this study. They are characterised by
minuscule coweights $\mu $ of the gauge symmetry G \textrm{\cite{J}}, and
have nice realisations in 4D Chern-Simons theory living on $M_{4}=\mathbb{R}%
^{2}\times \mathbb{CP}^{1}$ involving the compact complex holomorphic line $%
\mathbb{CP}^{1}$. In ref.\textrm{\cite{1F}}, the tH$_{\mathrm{\gamma }_{%
{\small z}}}^{\mu }$ lines\textrm{\ }were implemented\textrm{\ }in the 4D CS
theory through their coupling with a Wilson line W$_{\xi _{w}}^{q_{e}}$ in
some representation $\boldsymbol{R}$ of G. This tH$_{\mathrm{\gamma }_{%
{\small z}}}^{\mu }$-W$_{\xi _{w}}^{q_{e}}$ coupling is described by their
crossing as depicted by the Figure \textbf{\ref{Lop}}-a; see also \textrm{%
\cite{lax}}\textbf{. }They are modeled by a holomorphic matrix $\mathcal{L}%
\left( z\right) $ valued in the algebra $\mathfrak{A}$ of functions on the
phase space of tH$_{\mathrm{\gamma }_{{\small z}}}^{\mu };$ i.e: $\mathcal{L}%
\left( z\right) \in \mathfrak{A}\otimes End\left( \boldsymbol{R}\right) $.
Following \textrm{\cite{1F}}, minuscule 't Hooft lines, carrying a minuscule
magnetic charge $\mu ,$ can be remarkably described in 4D CS on $\mathbb{R}%
^{2}\times \mathbb{CP}^{1}$ with simply connected group G. Here, the real $%
\left( x,y\right) $ are local coordinates of $\mathbb{R}^{2},$ and the
complex $z=Z/Z^{\prime }$ is a local variable in $\mathbb{CP}^{1}$ with
homogeneous coordinates $\left( Z_{1},Z_{2}\right) \neq \left( 0,0\right) $.
In this 4D topological theory with field action $\int_{M_{4}}dz\wedge \Omega
_{3}\left( A\right) $ where $\Omega _{3}$ is the CS 3-form, the 't Hooft
lines are thought of as line defects in $M_{4}$ wrapping a real curve $%
\mathrm{\gamma }_{{\small z}}$ in $\mathbb{R}^{2}.$ It is convenient to take 
$\mathrm{\gamma }_{{\small z}}$ as the horizontal x-axis $\left \{
y=0,z=0\right \} $ which is interesting to imagine as given by $\mathrm{%
\gamma }_{{\small 0}}=\mathcal{O}_{-}^{{\small 0}}\cap \mathcal{O}_{+}^{%
{\small 0}}$ with $\mathbb{R}^{2}=\mathcal{O}_{-}^{{\small 0}}\cup \mathcal{O%
}_{+}^{{\small 0}};$ that is the intersection of two 2D patches $\mathcal{O}%
_{+}^{{\small z}}$ and $\mathcal{O}_{-}^{{\small z}}$ in $\mathbb{R}%
^{2}\times \mathbb{CP}^{1}$ like%
\begin{equation}
\mathcal{O}_{-}^{{\small 0}}=\left \{ \left( x,y;z\right) \text{ \ }|\text{
\ }\left. 
\begin{array}{c}
y\leq 0 \\ 
z=0%
\end{array}%
\right. \right \} ,\qquad \mathcal{O}_{+}^{{\small 0}}=\left \{ \left(
x,y;z\right) \text{ \ }|\text{ \ }\left. 
\begin{array}{c}
y\geq 0 \\ 
z=0%
\end{array}%
\right. \right \}  \label{pm}
\end{equation}%
These 't Hooft line defects can be also viewed as a pair tH$_{\mathrm{\gamma 
}_{{\small z}}}^{\pm \mu }$ living at the ends of a Dirac string at $y=0$
stretched between $z=0$ and $z=\infty .$ In this picture, the line defect tH$%
_{\mathrm{\gamma }_{{\small z}}}^{+\mu }$ at $z=0$ has a magnetic charge $%
+\mu $ and the line tH$_{\mathrm{\gamma }_{{\small z}}}^{-\mu }$ at infinity
has a magnetic charge $-\mu .$ Other interesting representations of tH$_{%
\mathrm{\gamma }_{{\small z}}}^{\pm \mu }$ in integrable field theory and
spin chains can be found in \textrm{\cite{1F} and refs therein}. For ADE
gauge symmetries G with generic rank $r$ given by simply connected groups
like the exceptional E$_{6}$ and E$_{7}$ we are considering in this study,
the G- bundle on the complex projective line is trivial. This remarkable
property is because of the trivial behavior of the gauge potential at $%
z=\infty $ and which extends to all points in $\mathbb{CP}^{1}.$ This
triviality feature on $\mathbb{CP}^{1}$ \textrm{applies} to the 2D patches $%
\mathcal{O}_{-}^{{\small z}}$ and $\mathcal{O}_{+}^{{\small z}}$ of eq(\ref%
{pm}) with $z\sim 0$; and has been used by Costello- Gaiotto- Yagi to
propose a classical gauge invariant observable $\mathcal{L}\left( z\right) $
given by the path ordered quantity $P\exp \int_{y}A\left( z\right) $ to
measure the parallel transport of the gauge field from the patch $\mathcal{O}%
_{y\leq 0}^{{\small z}}$ to the patch $\mathcal{O}_{y\geq 0}^{{\small z}}.$
It is a function of z; and belongs to the loop group $G\left( \left(
z\right) \right) $ of analytic functions valued in G. The CGY observable has
interesting properties described in \textrm{\cite{1F}}; in particular the%
\textrm{\ following ones that are useful for this study}: $\left( \mathbf{1}%
\right) $ As noticed before, the $\mathcal{L}\left( z\right) $ describes the
crossing of tH$_{\mathrm{\gamma }_{{\small z}}}^{\mu }$ with a Wilson line
as shown by the Figure \textbf{\ref{Lop}}. It plays a quite similar role as
the R-matrix of Wilson lines; and is also\emph{\ }interpreted in terms of
the \textrm{Lax operator of 2D integrable systems with spectral parameter z 
\cite{lax,L2}. }$\left( \mathbf{2}\right) $ the $\mathcal{L}\left( z\right) $
has poles and zeroes at $z=0$ and $z=\infty $ arising from 't Hooft lines tH$%
_{\mathrm{\gamma }_{{\small z}}}^{\pm \mu }$ at these particular points in $%
\mathbb{CP}^{1}.$ Near the singularity at $z=0$, the CGY observable can be
factorised like $\mathcal{L}\left( z\right) =A\left( z\right) z^{\mu
}B\left( z\right) $ where $A\left( z\right) $ and $B\left( z\right) $ are
regular functions at $z=0.$ The operator $z^{\mu }$ carries the Dirac
monopole singularity with minuscule coweight action given by the adjoint
form of $\mu $ to be described later. A quite similar factorisation of $%
\mathcal{L}\left( z\right) $ exists near the singularity at $z=\infty $. It
reads also as $\tilde{A}\left( z\right) z^{\mu }\tilde{B}\left( z\right) $;
but with $\tilde{A}\left( z\right) $ and $\tilde{B}\left( z\right) $ going
to identity for z approaching infinity. $\left( \mathbf{3}\right) $ \textrm{%
By imposing regularity conditions in bulk and \emph{boundary};} the $%
\mathcal{L}\left( z\right) $ can be brought to the following interesting form%
\begin{equation}
\mathcal{L}^{\left( \mu \right) }\left( z\right) =e^{X}z^{\mu }e^{Y}
\label{xy}
\end{equation}%
where $X$ and $Y$ are globally defined on $\mathbb{CP}^{1}$; i.e independent
of z. In this factorisation, the $X$ and $Y$ are matrix operators valued in
the nilpotent subalgebras \textbf{n}$_{+}$ and \textbf{n}$_{-}$ of the Levi-
decomposition with respect to $\mu $ of the Lie algebra $g$ of the gauge
symmetry G of the 4D CS theory. Notice that given a Lie algebra $g$, one may
have different minuscule coweights $\mu _{a}$ and consequently various tH$_{%
\mathrm{\gamma }_{{\small z}}}^{\mu _{a}}$ and different CGY observables $%
\mathcal{L}^{\left( \mu _{a}\right) }$. As an illustration, we describe
rapidly below the instructive example of 4D Chern-Simons theory with gauge
symmetry $G=SL\left( N\right) .$ The gauge group of this family has $N-1$
minuscule coweights $\mu _{a}$; and \textrm{so} $N-1$ types of minuscule 't
Hooft lines tH$_{\gamma .}^{\mu _{a}}.$ This property can be explicitly
formulated by using the canonical vector basis $\left \{ e_{i}=\left \vert
i\right \rangle \right \} $ of the ambient space $\mathbb{R}^{N}$ to express
the content of the root system $\Phi _{SL\left( N\right) }$ of SL$\left(
N\right) $ and its $N-1$ minuscule coweights $\mu _{1},...,\mu _{N-1}$.
Recall that the Lie algebra $sl_{N}$ underlying the SL$\left( N\right) $
gauge symmetry has $N\left( N-1\right) $ roots $\alpha =n_{a}\alpha _{a}$
generated by N-1 simple roots $\alpha _{a}=e_{a}-e_{a+1}$. It also has N-1
minuscule coweights obeying $\mu _{a}.\alpha _{b}=\delta _{ab}$ and
expressed in terms of the $e_{{\small i}}$'s like $\frac{{\small N-a}}{%
{\small N}}({\small e}_{{\small 1}}{\small +...+e}_{{\small a}}){\small -}%
\frac{{\small a}}{{\small N}}({\small e}_{{\small a+1}}{\small +...+e}_{%
{\small N}}).$ The \textrm{adjoint} form of the minuscule coweights used in $%
\mathcal{L}^{\left( \mu _{a}\right) }$ reads as $\mu _{a}^{i}\left \vert
i\right \rangle \left \langle i\right \vert $ with $\mu _{a}^{i}=1-a/N$ for $%
1\leq i\leq a$ and $-a/N$ for $a+1\leq i\leq N$. It happens that for the SL$%
\left( N\right) $ family, the matrix operators $X$ and $Y$ in (\ref{xy})
obey the nilpotency relations $X^{2}=Y^{2}=0$; then the $\mathcal{L}$%
-operator for SL$\left( N\right) $ reduces to the following particular form%
\begin{equation}
\mathcal{L}_{sl_{N}}\left( z\right) =\left( I+X\right) z^{\mu }\left(
I+Y\right)
\end{equation}%
Notice that in the above $\mathcal{L}_{sl_{N}}\left( z\right) $, the higher
monomial in the nilpotent matrix operators is $Xz^{\mu }Y.$ This is a
specific property of $sl_{N}.$ Later \textrm{on, we will} study the
extension of this construction to \textrm{the} 4D Chern-Simons gauge
theories with exceptional E$_{6}$ and E$_{7}$ gauge symmetries. \textrm{Then}%
, we will derive new aspects regarding the structure of the topological $%
\mathcal{L}_{E_{6}}$ and $\mathcal{L}_{E_{7}}$.

\subsection{Phase space of 't Hooft lines}

\begin{figure}[tbph]
\begin{center}
\includegraphics[width=12cm]{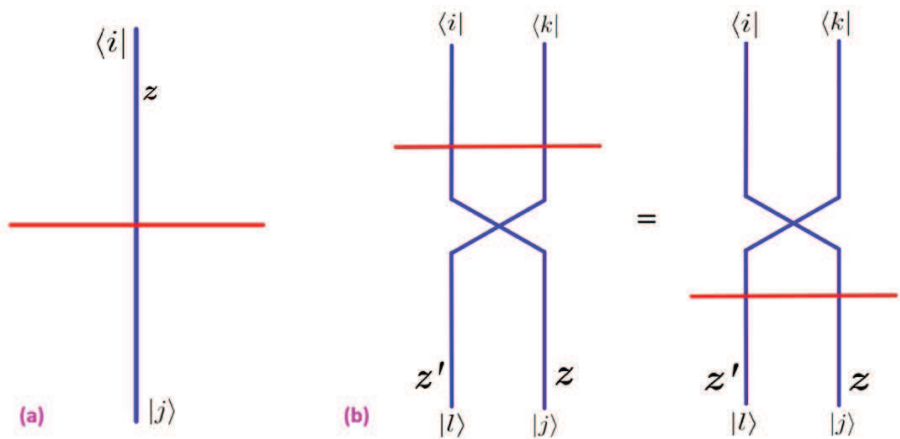}
\end{center}
\par
\vspace{-0.5cm}
\caption{$\left( \mathbf{a}\right) $ The operator $\mathcal{L}\left(
z\right) $ encoding the coupling between a 't Hooft line at z=0 (in red) and
a Wilson line at z (in blue) with incoming $\left \langle i\right \vert $
and out going $\left \vert j\right \rangle $ states. $\left( \mathbf{b}%
\right) $ RLL relations encoding the commutation relations between two
L-operators at z and z'.}
\label{Lop}
\end{figure}

Here, we describe the classical phase space $\mathcal{E}_{ph}\left[ L\left(
z\right) \right] $ of the minuscule 't Hooft lines in 4D CS theory and its
parametrisation using Darboux coordinates. An interesting way to deal with
the properties of $\mathcal{E}_{ph}$ is to consider the CGY observable $%
\mathcal{L}\left( z\right) $ and use the graphic representation depicted by
the Figures \textbf{\ref{Lop}}. In this representation, we think of the CGY
observable as a matrix operator $\left \langle i|\mathcal{L}\left( z\right)
|j\right \rangle $ describing the crossing of a tH$_{\gamma _{0}}^{\mu }$
with a Wilson line W$_{\xi _{z}}^{q_{e}}$ on which a set of $\left \vert
i\right \rangle $- states propagate. In this picture, the 't Hooft line is
materialized by the horizontal x-axis of the plane $\mathbb{R}^{2}$ with
spectral parameter $z=0$; and the Wilson line is given by the vertical
y-axis with a generic $z$ in $\mathbb{CP}^{1}$. The charge of the tH$%
_{\gamma _{0}}^{\mu }$ is given by the minuscule coweight $\mu $ of the
gauge symmetry G and the Wilson line is characterised by some representation 
$\boldsymbol{R}$ of G with incoming states $\left \vert i\right \rangle $
and outgoing $\left \vert j\right \rangle $. 
As such, the $\mathcal{L}\left( z\right) $ combines topological data from
the tH$_{\gamma _{0}}^{\mu }$ and the Wilson W$_{\xi _{z}}^{q_{e}}$ encoded
in the $L_{j}^{i}\left( z\right) $ matrix entries of $\mathcal{L}$. In this
view, the symplectic structure of two operators $L_{j}^{r}\left( z\right) $
and $L_{l}^{s}\left( z^{\prime }\right) $ is given by the RLL relations
shown \textrm{in} the Figure \textbf{\ref{Lop}}-(b). These relations are due
to the topological invariance of the CS theory and read explicitly as
follows \textrm{\cite{1B},}%
\begin{equation}
R_{rs}^{ik}\left( z-z^{\prime }\right) L_{j}^{r}\left( z\right)
L_{l}^{s}\left( z^{\prime }\right) =L_{r}^{i}\left( z^{\prime }\right)
L_{s}^{k}\left( z\right) R_{jl}^{rs}\left( z-z^{\prime }\right)  \label{rll}
\end{equation}%
where the tensor $R_{rs}^{ik}\left( z-z^{\prime }\right) $ is the usual
R-operator used in the study of Yang- Baxter equation (YBE). Though looking
cumbersome, the RLL relations (\ref{rll}) were shown to be equivalent, 
\textrm{at the leading order in the }$\hbar $\textrm{-expansion of the R
matrix ---see (\ref{2}) given below---,} to the usual Poisson bracket $%
\left
\{ b^{\alpha },c_{\beta }\right \} _{PB}=\delta _{\beta }^{\alpha }$
of symplectic geometry with Darboux coordinates $\left( b,c\right) $
imagined in this limit as classical oscillators. The equivalence between $%
L_{j}^{i}$ and $\left( b^{\alpha },c_{\beta }\right) $ is established using
the Levi- decomposition of the Lie algebra $g$ of the gauge symmetry G.
Indeed, given a minuscule coweight $\mathbf{\mu }$ of $g$, we have the
following Levi-decomposition \textrm{\cite{Levi}} 
\begin{equation}
g=\boldsymbol{l}_{\mu }\oplus \boldsymbol{n}_{+}\oplus \boldsymbol{n}_{-}=%
\boldsymbol{p}\oplus \boldsymbol{n}_{-}  \label{eq}
\end{equation}%
where $\boldsymbol{l}_{\mu }$ is the Levi- factor and the $\boldsymbol{n}%
_{\pm }$ are nilpotent subalgebras of $g$\texttt{.} In the second equality
of (\ref{eq}), we have used the short splitting$\ p=l_{\mu }\oplus n_{+}$
with $p$ standing for a parabolic subalgebra of $g$. By using Dirac
singularity at $z=0$ and $z=\infty $ and following \textrm{\cite{1F}}, we
can first factorise the $L\left( z\right) $ operator as in eq(\ref{xy})
where $X$ and $Y$ are nilpotent operators respectively valued in the
subalgebras $\boldsymbol{n}_{+}$ and $\boldsymbol{n}_{-}.$ These nilpotent
operators do not depend on the spectral parameter z; they are globally
defined on $\mathbb{CP}^{1}$. Moreover denoting by $X_{\alpha }$ the
generators of $\boldsymbol{n}_{+}$ and by $Y^{\beta }$ the generators of $%
\boldsymbol{n}_{-},$ then substituting the expansions $X=b^{\alpha
}X_{\alpha }$ and $Y=c_{\beta }Y^{\beta }$ into (\ref{rll}) with (\ref{xy})
and replacing $R_{rs}^{ik}$ by its expression in terms of the double
Casimirs, we end up with the wanted Poisson bracket 
\begin{equation}
\left \{ b^{\alpha },c_{\beta }\right \} _{PB}=\delta _{\beta }^{\alpha }
\label{cb}
\end{equation}%
showing that $b^{\alpha }$ and $c_{\beta }$ are indeed Darboux-like complex
coordinates of the classical phase space $\mathcal{E}_{ph}\left[ L\left(
z\right) \right] $ of the minuscule 't Hooft lines in 4D CS gauge theory. 
\textrm{Here, we used the expression of\emph{\ }the R-matrix as a rational
solution of the YBE at the leading order in }$\hbar $\textrm{\
(semi-classical) \cite{1A,2A}}%
\begin{equation}
R_{jl}^{ik}(z)=\delta _{j}^{i}\delta _{l}^{k}+\frac{\hbar }{z}c_{jl}^{ik}+O(h%
{{}^2}%
)  \label{2}
\end{equation}%
\textrm{where }$c_{jl}^{ik}$\textrm{\ stands for the double Casimir of the
gauge symmetry; its value for sl}$_{N}$\textrm{\ is }$\delta _{l}^{i}\delta
_{j}^{k}.$\textrm{\ Notice as well that at the }quantum level, the $%
b^{\alpha }$ and $c_{\beta }$ are promoted to operators and eq(\ref{cb}) is
replaced by the commutator $[\hat{c}_{\beta },\hat{b}^{\alpha }]\sim \hbar
\delta _{\beta }^{\alpha }$ \textrm{where} $\hat{c}_{\beta }$ and $\hat{b}%
^{\alpha }$ respectively interpreted as annihilation and creation operators.
To get more insight into this promotion, \textrm{note} that by substituting $%
\hat{X}=\hat{b}^{\alpha }X_{\alpha }$ and $\hat{Y}=\hat{c}_{\beta }Y^{\beta
} $ into (\ref{xy}), we obtain $e^{\hat{b}^{\alpha }X_{\alpha }}z^{\mu }e^{%
\hat{c}_{\beta }Y^{\beta }}.$ In this relation, the creators $e^{\hat{b}%
^{\alpha }X_{\alpha }}$ are put on the left and the annihilators $e^{\hat{c}%
_{\beta }Y^{\beta }}$ are put on the right; thus indicating that quantum
mechanically speaking, the Lax operator given by eq(\ref{xy}) is non
ambiguous as it is normal ordered; that is $\mathcal{\hat{L}}^{\left( \mu
\right) }=:e^{\hat{X}}z^{\mu }e^{\hat{Y}}:$. Notice \textrm{moreover} that
for the case of the Levi- decomposition $sl_{N}\rightarrow \boldsymbol{l}%
_{\mu _{1}}\oplus \boldsymbol{n}_{+}\oplus \boldsymbol{n}_{-}$ with $%
\boldsymbol{l}_{\mu _{1}}=sl_{N-1}\oplus \mathbb{C}\mu _{1}$ and $%
\boldsymbol{n}_{\pm }=\mathbf{F}_{\pm }$ with $sl_{N-1}$ fundamentals $F=N-1$
and its dual, the classical $\mathcal{L}$- operator reads as follows :%
\begin{equation}
\mathcal{L}_{sl_{N}}=\left( 
\begin{array}{cc}
z^{\frac{N-1}{N}}+\mathbf{b}^{T}\mathbf{c} & \mathbf{b}^{T} \\ 
\mathbf{c} & z^{-\frac{1}{N}}I_{N-1}%
\end{array}%
\right)
\end{equation}%
with $\mathbf{b}^{T}=\left( b_{1},...,b_{N-1}\right) $ and $\mathbf{c}%
=\left( c_{1},...,c_{N-1}\right) ^{T}.$ Having revisited useful aspects on
the CGY observable in 4D CS theory on $\Sigma \times \mathbb{CP}^{1}$; we
turn now to \textrm{represent} our contribution by focussing first on the 'tH%
$_{\mathrm{\gamma }}^{\mu }$ in the E$_{6}$ CS gauge theory and then on the
case of the E$_{7}$ theory.

\section{Exceptional E$_{6}$ minuscule 't Hooft line}

In this section, we study the exceptional minuscule 't Hooft line operators
living in \textrm{the} 4D CS theory with gauge symmetry E$_{6}$. From the
analysis of the previous section, the phase space $\mathcal{E}_{ph}^{E_{6}}$
of the operator $\mathcal{L}_{E_{6}}$ is determined by using the Levi-
decomposition of the gauge symmetry with respect to a minuscule coweight $%
\mathbf{\mu }_{E_{6}}$ of the \textrm{Lie algebra }$\mathrm{e}_{6}$\textrm{\
of the Lie group E}$_{6}.$ Before constructing $\mathcal{L}_{E_{6}}\left(
z\right) $, let us start by giving some useful tools regarding E$_{6},$ its
Levi- decomposition and the splitting of its fundamental representation 
\textbf{27}.

\subsection{Minuscule coweights and Levi subalgebra of E$_{6}$}

\begin{figure}[tbph]
\begin{center}
\includegraphics[width=8cm]{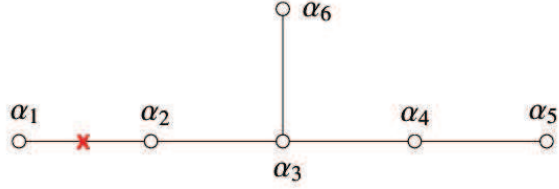}
\end{center}
\par
\vspace{-0.5cm}
\caption{Dynkin Diagram of E6 having six nodes labeled by the simple roots $%
\protect \alpha _{i}$. The cross $\left( \times \right) $ indicates the roots
used in the Levi decomposition with Levi subalgebra $so(10)\oplus so(2).$}
\label{E6}
\end{figure}

The exceptional Lie algebra $\boldsymbol{e}_{6}$ has six Cartan type
generators H$_{\alpha _{i}}$ and 72 step operators $\mathcal{Z}_{\pm \alpha
} $ labeled by the $\pm \alpha $ roots of E$_{6}$ with length $\alpha
^{2}=2. $ The 72 roots of the root system $\Phi _{E_{6}}$ are generated by
six simple roots $\left \{ \alpha _{i}\right \} _{1\leq i\leq 6}$ with
intersection K$_{ij}$ given by the Cartan matrix K$_{E_{6}}$. We realise
these roots in $\mathbb{R}^{8}$ as follows 
\begin{equation}
\alpha _{1}=\frac{1}{2}\left( \epsilon _{1}-\epsilon _{2}-\epsilon
_{3}-\epsilon _{4}-\epsilon _{5}-\epsilon _{6}-\epsilon _{7}+\epsilon
_{8}\right)  \label{al1}
\end{equation}%
\emph{and }$\alpha _{i}=\epsilon _{i}-\epsilon _{i-1}$\emph{\ for }$i\neq
1,6 $ as well as $\alpha _{6}=\epsilon _{1}+\epsilon _{2}.$ The 72 roots $%
\alpha $ of E$_{6}$ can be organised into two subsystems. $\left( \mathbf{1}%
\right) $ a subset of 40 positive $\pm \left( \epsilon _{i}\pm \epsilon
_{j}\right) $ with $1\leq j<i\leq 5$ $;$ and $\left( \mathbf{2}\right) $ a
subset of 32 roots\ $\pm \frac{1}{2}\left( q_{i}\epsilon _{i}-\epsilon
_{6}-\epsilon _{7}+\epsilon _{8}\right) $ where the five $q_{i}$ take $\pm 1$
with $\Pi _{i=1}^{5}q_{i}=1.$ Concerning the representations of the Lie
algebra of E$_{6},$ they can be built out of its six fundamental
representations. Here, we will be particularly interested into 78$_{0},$
associated with the simple root $\alpha _{6}$ as depicted by the the Figure 
\textbf{\ref{E6}}, and into the 27$_{\pm }$ associated with $\alpha _{1}$
and $\alpha _{5}.$ The E$_{6}$ has two minuscule coweights $\mu _{1}$ and $%
\mu _{5}$ dual to $\alpha _{1}$ and $\alpha _{5}$; they \textrm{respectively 
}correspond to the fundamentals $27_{+}$ and $27_{-}$. Taking as a minuscule 
$\mu $ for our E$_{6}$ gauge theory the $\mu _{1}$ coweight, it follows that
the Levi- subalgebra $\boldsymbol{l}_{\mu }$ of the exceptional E$_{6}$ is
given by $so(2)\oplus so(10).$ Thus, the 40 roots of $so(10)$ is a subset of
the root system of E$_{6}$; it is easily read from the E$_{6}$ system by
omitting the 32 roots containing the spinorial-like root $\alpha _{1}$. In
this view, the simple roots of the so$\left( 10\right) $ subalgebra of $%
\boldsymbol{e}_{6}$ are given the five $\alpha _{2},\alpha _{3},\alpha
_{4},\alpha _{5},\alpha _{6}$ and the Levi- decomposition $\boldsymbol{e}%
_{6}=\boldsymbol{l}_{\mu }\oplus \boldsymbol{n}_{+}\oplus \boldsymbol{n}_{-}$
reads as follows%
\begin{equation}
e_{6}\rightarrow so(2)\oplus so(10)\oplus \boldsymbol{16}_{+}\oplus 
\boldsymbol{16}_{-}  \label{e6d}
\end{equation}%
This splitting distributes the 78 dimensions of $\boldsymbol{e}_{6}$ like $%
1+45+16_{+}+16_{-}$. Eq(\ref{e6d}) can be also read at the level of the
Dynkin diagram of $\boldsymbol{e}_{6}$ given by the Figure \textbf{\ref{E6}}%
. \textrm{By cutting the node} $\alpha _{1}$, associated with the minuscule
coweight $\mu _{1}$, we recover the Dynkin diagram of $so(10)$ and its
spinor representations $16_{+}$ and $16_{-}$ charged under so$\left(
2\right) $. Notice that the 36+36 step operators of $\boldsymbol{e}_{6}$ are
split in the Levi- decomposition as 20+20 step operators $Z_{\pm \alpha }$
generating so$\left( 10\right) ,$ 16 step operators $X_{+\beta }$ generating
the nilpotent subalgebra $\boldsymbol{16}_{+}$ and 16 other step operators $%
X_{-\beta }=Y^{\beta }$ generating $\boldsymbol{16}_{-}$. 
Notice also that under the Levi- decomposition, representations $\boldsymbol{%
R}_{E_{6}}$ of the exceptional symmetry E$_{6}$ reduce as direct sums $\sum
\left( \boldsymbol{R}_{l}^{so_{10}},\boldsymbol{R}_{l}^{so_{2}}\right) $ of
representations of $so(10)\oplus so(2).$ For the example of the fundamental
of E$_{6}$, we have the following splitting \textrm{\cite{br}}, 
\begin{equation}
\mathbf{27}=(\mathbf{1},-\frac{4}{3})+(\mathbf{10},+\frac{2}{3})+(\mathbf{16}%
,-\frac{1}{3})  \label{27}
\end{equation}

\subsection{Minuscule CGY observable $\mathcal{L}_{E_{6}}$}

\begin{figure}[tbph]
\begin{center}
\includegraphics[width=8cm]{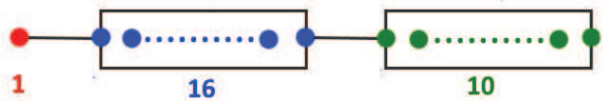}
\end{center}
\par
\vspace{-0.5cm}
\caption{A formal graphic illustrating Levi-decomposition of the
representation 27 of E$_{6}$ in terms of representations of SO$\left(
10\right) .$ Here, we have $27=1+16+10$.}
\label{270}
\end{figure}

To construct the 't Hoof line operator of the exceptional E$_{6}$ CS theory,
notice that $\mathcal{L}_{E_{6}}^{\left( \mu \right) }$ is characterised by
the \textrm{representation }$\boldsymbol{R}$ of the Wilson line and by the
minuscule coweight $\mu $ with Levi- quantum numbers as in (\ref{27}). This $%
\mathcal{L}_{E_{6}}^{\left( \mu \right) }$ operator is given by the Levi-
factorisation $e^{X}z^{\mathbf{\mu }}e^{Y}$ with $X=\sum_{\beta
=1}^{16}b^{^{\beta }}X_{\beta }$ and $Y=\sum_{\beta =1}^{16}c_{\beta
}Y^{\beta }.$ In these expansions, the $b^{\beta }$ and $c_{\beta }$ stand
for the 16+16 Darboux coordinates satisfying the Poisson bracket $\left \{
b^{\gamma },c_{\beta }\right \} =\delta _{\beta }^{\gamma }$ and $X_{\beta }$
and $Y^{\beta }$ are the generators of the nilpotent subalgebras \textbf{16}$%
_{\pm }.$ The charge operator $\mathbf{\mu }$ associated with the minuscule
coweight is given by 
\begin{equation}
\mathbf{\mu }=-\frac{4}{3}\varrho _{\underline{\mathbf{1}}}+\frac{2}{3}%
\varrho _{\underline{\mathbf{10}}}-\frac{1}{3}\varrho _{\underline{\mathbf{16%
}}}  \label{mu}
\end{equation}%
where $\varrho _{\underline{\mathbf{1}}},$ $\varrho _{\underline{\mathbf{10}}%
}$ and $\varrho _{\underline{\mathbf{16}}}$ are projectors on the $%
so(10)\oplus so(2)$ representation spaces of (\ref{27}). \textrm{If we
choose to denote} the 27 states of the fundamental representation of E$_{6}$
by the basis vector kets $\left \vert \xi \right \rangle $ with $\xi
=0,1,...26 $ as formally depicted by the Figure \textbf{\ref{270}}, 
then the projectors are given by $\varrho _{\underline{\mathbf{1}}%
}=\left
\vert 0\right \rangle \left \langle 0\right \vert $ and $\varrho _{%
\underline{\mathbf{10}}}=\sum_{l=1}^{10}\left \vert v_{l}\right \rangle
\left \langle v^{l}\right \vert $ as well as $\varrho _{\underline{\mathbf{16%
}}}=\sum_{\beta =1}^{16}\left \vert s_{\beta }\right \rangle \left \langle
s^{\beta }\right \vert .$ Using the basis state kets $\left \vert
0\right
\rangle ,$ $\left \vert v_{l}\right \rangle $ and $\left \vert
s_{\beta }\right \rangle $ satisfying \textrm{the} orthogonality properties $%
\left
\langle 0|v_{l}\right \rangle =\left \langle 0|s_{\beta
}\right
\rangle =\left \langle v_{l}|s_{\beta }\right \rangle =0$, we can
explicitly realise the generators $X_{\beta }$ and $Y^{\beta }$ in the phase
space of the E$_{6} $ 't Hooft line operator. We find%
\begin{equation}
X_{\beta }=\left( \Gamma ^{i}\right) _{\beta \gamma }\left \vert v_{i}\right
\rangle \left \langle s^{\gamma }\right \vert +\left \vert s_{\beta }\right
\rangle \left \langle 0\right \vert ,\quad Y^{\beta }=\left \vert 0\right
\rangle \left \langle s^{\beta }\right \vert +\left( \Gamma _{i}\right)
^{\beta \gamma }\left \vert s_{\gamma }\right \rangle \left \langle
v^{i}\right \vert  \label{rxy}
\end{equation}%
where the $\Gamma _{i}$'s are the ten-dimensional Gamma matrices satisfying
the usual Clifford algebra $\Gamma _{i}\Gamma _{j}+\Gamma _{j}\Gamma
_{i}=2\delta _{ij}$. Notice that \textrm{one can also construct operators of 
}$so(10)\oplus so(2)$\textrm{\ from these }$\left \vert 0\right \rangle ,$ $%
\left \vert v_{l}\right \rangle $ and $\left \vert s_{\beta }\right \rangle $%
\textrm{\ states. An interesting operator is the generator of }$so(2)$ which
is nothing but (\ref{mu}); it acts on $X_{\beta }$ and $Y^{\beta }$ like $%
\left[ \mathbf{\mu },X_{\beta }\right] =X_{\beta }$ and $\left[ \mu
,Y^{\beta }\right] =-Y^{\beta }$ as required by the Levi-decomposition. As a
direct check, we \textrm{can calculate the first commutator as} 
\begin{equation}
\left[ \mathbf{\mu },X_{\beta }\right] =(-\frac{1}{3}+\frac{4}{3})\left
\vert s_{\beta }\right \rangle \left \langle 0\right \vert +(\frac{2}{3}+%
\frac{1}{3})\left( \Gamma ^{i}\right) _{\beta \gamma }\left \vert
v_{i}\right \rangle \left \langle s^{\gamma }\right \vert
\end{equation}%
which is equal to $X_{\beta }$. Using the realisation (\ref{rxy}), we can 
\textrm{work out }explicit calculations regarding the CGY observable $%
\mathcal{L}_{E_{6}}$. From (\ref{rxy}), we \textrm{deduce} the action of the 
$X_{\beta }$ and $Y^{\beta }$ generators of the 16$_{+}$ and 16$_{-}$ blocks
in (\ref{e6d}) on the states $\left \vert 0\right \rangle ,$ $\left \vert
v^{i}\right \rangle $ and $\left \vert s^{\beta }\right \rangle $. As such,
we can write their explicit matrix realisations$\left \langle A|X_{\beta
}|B\right \rangle $ and $\left \langle A|Y^{\beta }|B\right \rangle $ with
labels $A,B=0,l,\beta $. We can also compute the powers of the operators $%
X=b^{\beta }X_{\beta }$ and $Y=c_{\beta }Y^{\beta }$ involved in the
calculation of $e^{X}z^{\mu }e^{Y}$. We find that $X^{3}=Y^{3}=0$ and 
\begin{equation}
X^{2}=2V^{i}\left \vert v_{i}\right \rangle \left \langle 0\right \vert
\qquad ,\qquad Y^{2}=2W_{i}\left \vert 0\right \rangle \left \langle
v^{i}\right \vert  \label{nl}
\end{equation}%
where we have set $V^{i}=\frac{1}{2}b^{\alpha }\left( \Gamma ^{i}\right)
_{\alpha \beta }b^{\beta }$ and $W_{i}=\frac{1}{2}c_{\alpha }\left( \Gamma
_{i}\right) ^{\alpha \beta }c_{\beta }$. The nilpotency feature of the X and
Y operators leads to the finite expansion $e^{X}=I+X+\frac{1}{2}X^{2}$ and
the same goes for $e^{Y}.$ Substituting (\ref{rxy}) and (\ref{nl}) into
these expansions, we obtain 
\begin{equation}
\begin{tabular}{lll}
$e^{X}$ & $=$ & $I+b^{\beta }\left( \Gamma _{\beta \gamma }^{i}\left \vert
v_{i}\right \rangle \left \langle s^{\gamma }\right \vert +\left \vert
s_{\beta }\right \rangle \left \langle 0\right \vert \right) +V^{i}\left
\vert v_{i}\right \rangle \left \langle 0\right \vert $ \\ 
$e^{Y}$ & $=$ & $I+c_{\beta }\left( \left \vert 0\right \rangle \left
\langle s^{\beta }\right \vert +\Gamma _{i}^{\beta \gamma }\left \vert
s_{\gamma }\right \rangle \left \langle v^{i}\right \vert \right)
+W_{i}\left \vert 0\right \rangle \left \langle v^{i}\right \vert $%
\end{tabular}
\label{zm}
\end{equation}%
reading in matrix notation M$_{B}^{A}=\left \langle \xi _{B}|M|\xi
^{A}\right \rangle $\textrm{\ in the} basis ordered like $\left \vert \xi
_{B}\right \rangle =\left \vert 0\right \rangle ,\left \vert
v_{j}\right
\rangle ,\left \vert s_{\beta }\right \rangle $ as follows%
\begin{equation}
\left( e^{X}\right) _{B}^{A}=\left( 
\begin{array}{ccc}
1 & 0 & 0 \\ 
V^{i} & \delta _{j}^{i} & B_{\beta }^{i} \\ 
b^{\alpha } & 0 & \delta _{\beta }^{\alpha }%
\end{array}%
\right) ,\qquad \left( z^{\mathbf{\mu }}e^{Y}\right) _{B}^{A}=\left( 
\begin{array}{ccc}
z^{-\frac{4}{3}} & z^{-\frac{4}{3}}W_{j} & z^{-\frac{4}{3}}c_{\beta } \\ 
0\mathrm{\ } & z^{\frac{2}{3}}\delta _{j}^{i} & 0 \\ 
0 & z^{-\frac{1}{3}}C_{j}^{\alpha } & z^{-\frac{1}{3}}\delta _{\beta
}^{\alpha }%
\end{array}%
\right)
\end{equation}%
where we have set $B_{\beta }^{i}=b^{\gamma }\Gamma _{\gamma \beta }^{i}$
and $C_{j}^{\alpha }=c_{\gamma }\Gamma _{j}^{\gamma \alpha }$. Substituting,
we end up with the expression of the CGY- observable given by%
\begin{equation}
\mathcal{L}_{E_{6}}=\left( 
\begin{array}{ccc}
z^{-\frac{4}{3}} & z^{-\frac{4}{3}}W_{j} & z^{-\frac{4}{3}}c_{\beta } \\ 
z^{-\frac{4}{3}}V^{i} & z^{\frac{2}{3}}\delta _{j}^{i}+z^{-\frac{4}{3}%
}V^{i}W_{j}+z^{-\frac{1}{3}}B_{\alpha }^{i}C_{j}^{\alpha } & z^{-\frac{4}{3}%
}V^{i}c_{\beta }+z^{-\frac{1}{3}}B_{\beta }^{i} \\ 
z^{-\frac{4}{3}}b^{\alpha } & z^{-\frac{4}{3}}b^{\alpha }W_{j}+z^{-\frac{1}{3%
}}C_{j}^{\alpha } & z^{-\frac{1}{3}}\delta _{\beta }^{\alpha }+z^{-\frac{4}{3%
}}b^{\alpha }c_{\beta }%
\end{array}%
\right)  \label{en}
\end{equation}%
\textrm{We end this subsection by mentioning that if instead of cutting the
simple root }$\alpha _{1}$ \textrm{in the Figure \textbf{\ref{E6}}, we omit
the simple root }$\alpha _{5}$, we follow \textrm{an analogous analysis to
the one represented} here. In the $\alpha _{5}$- dual description, \textrm{%
we obtain quite similar results as }$\mathcal{L}_{E_{6}}^{\mu _{1}}\left[ z%
\right] $\textrm{; but with replacing the spectral parameter z with }$w=%
\frac{1}{z};$ that is 
\begin{equation}
\mathcal{L}_{E_{6}}^{\mu _{5}}\left[ z\right] =\mathcal{L}_{E_{6}}^{\mu
_{1}}[1/z]
\end{equation}%
\textrm{\ This feature may be nicely viewed from the correspondence between
the simple roots }$\alpha _{5}/\alpha _{1}$\textrm{\ and the fundamental
coweights }$\mu _{5}/\mu _{1}$\textrm{. In this regard, recall that }$\mu
_{5}$\textrm{\ is the highest coweight of the fundamental representation }$%
27_{-}$\textrm{\ just like }$\mu _{1}$\textrm{\ is the highest coweight of
the fundamental representation }$27_{+}$\textrm{\ given by eq(\ref{27}).
Since the two coweights are related as }$\mu _{5}=-\mu _{1},$\textrm{\ it
follows that (\ref{mu}) gets modified as }$\bar{\mu}=\frac{4}{3}\varrho _{%
\mathbf{\bar{1}}}-\frac{2}{3}\varrho _{\overline{\mathbf{10}}}+\frac{1}{3}%
\varrho _{\overline{\mathbf{16}}}$\textrm{. }

\section{Quiver representation of $\mathcal{L}_{E_{6}}$}

\begin{figure}[tbph]
\begin{center}
\includegraphics[width=5cm]{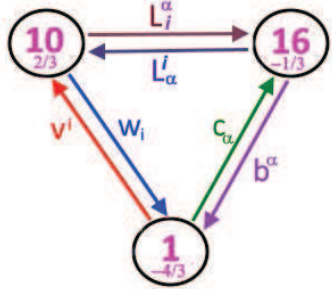}
\end{center}
\par
\vspace{-0.5cm}
\caption{$\mathcal{L}_{E_{6}}$ as a topological quiver with 3 nodes $N_{i}$
and $6$ links \textsc{l}$_{ij}$. The nodes are given by the self-dual $%
R_{i}\otimes \bar{R}_{i}$ and the links by bi-matter $R_{i}\otimes \bar{R}%
_{j}.$ In addition to SO$\left( 10\right) $ representations, the Darboux
coordinates $b^{\protect \alpha },$ $c_{\protect \alpha }$ carry an SO$\left(
2\right) $ charge $q=-1,$ \ $+1$. The fundamental vector like matter $V^{i}$
and $W_{i}$ carry $-2$ and $+2$.}
\label{so10}
\end{figure}

In this section, we develop a graphic representation of the CGY- operator $%
\mathcal{L}_{E_{6}}$ and use it to comment on the topological structure of
the $L_{{\small B}}^{{\small C}}$ entries of (\ref{en}). We denote this
graph as Q$_{E_{6}}$ and we refer to it as the topological gauge quiver
associated with the line operator $\mathcal{L}_{E_{6}}$. This denomination
is borrowed from supersymmetric quiver gauge theory with gauge symmetry $%
G=\Pi _{i}G_{i}$ where the nodes of the supersymmetric gauge quiver
represent the gauge factors $G_{i}$ and its links $R_{i\bar{j}}\sim
R_{i}\times \bar{R}_{j}$ describe the bi-fundamental matter in ($G_{i},G_{j}$%
) \textrm{\cite{bif,eh}}. Concerning the topological quiver Q$_{E_{6}}$ we
are interested in here, it is motivated amongst others (see below) by $%
\left( \mathbf{i}\right) $ the Levi- decomposition $N_{-}\times \boldsymbol{L%
}_{\mu }\times N_{+}$ of the gauge symmetry E$_{6},$ and $\left( \mathbf{ii}%
\right) $ the topological aspect of the 4D CS theory. \newline
A way to introduce the topological quiver Q$_{E_{6}}$ of the CGY operator $%
\mathcal{L}_{E_{6}}$ is to use the phase space bracket (\ref{rll}) solved by
eq(\ref{cb}). This last relation indicates that the Darboux coordinates $%
\left( b^{\beta },c_{\gamma }\right) $ are fundamental objects in dealing
with $\mathcal{L}_{E_{6}}$. As such, they play a quite similar role as the
(bi-) fundamental matter $R_{i\bar{j}}$ in supersymmetric quiver gauge
theories. So, it is interesting to use this property of $b^{\beta }$ and $%
c_{\gamma }$ to encode the internal structure of $\mathcal{L}_{E_{6}}$ into
a diagram Q$_{E_{6}}$ with three nodes N$_{1}$, N$_{2}$, N$_{3}$ related to
each other by 3+3 links \textsc{l}$_{ij}$ and \textsc{l}$_{ji}$ with $1\leq
i<j\leq 3$ as depicted by the Figure \textbf{\ref{so10}}. 
The nodes and the links of the quiver Q$_{E_{6}}$ are determined from the
matrix entries $L_{{\small B}}^{{\small C}}$ of the CGY- observable $%
\mathcal{L}_{E_{6}}$ and have an interpretation in terms of representations
of the Levi- subalgebra of $\boldsymbol{e}_{6}$. Indeed, from the
decomposition $e_{6}\rightarrow 16_{-}\oplus \boldsymbol{l}_{\mu }\oplus
16_{+}$ with $\boldsymbol{l}_{\mu }=so(2)\oplus so(10)$ and the expansions $%
X=b^{\beta }X_{\beta }$ and $Y=c_{\beta }Y^{\beta }$, we see that the
Darboux coordinates transform under $\boldsymbol{l}_{\mu }$ like $b^{\beta
}\sim 16_{-}$ and $c_{\gamma }\sim 16_{+}$. So, we can think of the $%
b^{\beta }$ and the $c_{\gamma }$ in terms of (topological) bi-fundamental
matter of $SO(2)\times SO(10)$ as described below. First, we observe that
the operator $\mathcal{L}_{E_{6}}=e^{X}z^{\mathbf{\mu }}e^{Y}$ is a $%
27\times 27^{t}$ matrix with entries $L_{{\small B}}^{{\small C}}$ given by
eq(\ref{en}). Second, we split the $L_{{\small B}}^{{\small C}}$ like%
\begin{equation}
\mathcal{L}_{E_{6}}=\left( 
\begin{array}{ccc}
L_{{\small 0}}^{{\small 0}} & L_{{\small 0}}^{{\small i}} & L_{{\small 0}}^{%
{\small \gamma }} \\ 
L_{{\small j}}^{{\small 0}} & L_{{\small j}}^{{\small i}} & L_{{\small j}}^{%
{\small \gamma }} \\ 
L_{{\small \beta }}^{{\small 0}} & L_{{\small \beta }}^{{\small i}} & L_{%
{\small \beta }}^{{\small \gamma }}%
\end{array}%
\right)  \label{ml}
\end{equation}%
and look for the algebraic structure of the sub-block entries. Obviously the
entries $L_{{\small B}}^{{\small C}}$ are given by particular polynomial
functions $f(b,c)$ of the 16+16 Darboux coordinates $b^{\beta }$ and $%
c_{\gamma }$. In writing (\ref{ml}), we have decomposed the indices B and C
in terms of three labels like $B=\left( 0,j,\beta \right) $ and $C=\left(
0,i,\gamma \right) $ corresponding to the decomposition $%
27=1_{-4/3}+10_{+2/3}+16_{-1/3}$ of the fundamental representation of E$_{6}$%
. As such, the operator $\mathcal{L}_{E_{6}}$ has, generally speaking, $729$
components $L_{{\small B}}^{{\small C}}$ that can be decomposed as $%
1+78+650. $ In terms of tensor products of SO$\left( 10\right) $
representations, we also have%
\begin{equation}
\begin{tabular}{lllllllllll}
$27\times \overline{27}=$ & $1\times \overline{1}$ & $\oplus $ & $1\times 
\overline{10}$ & $\oplus $ & $1\times \overline{16}$ & $\oplus $ & $16\times 
\overline{1}$ & $\oplus $ & $16\times \overline{10}$ & $\oplus $ \\ 
& $10\times \overline{1}$ & $\oplus $ & $10\times \overline{10}$ & $\oplus $
& $10\times \overline{16}$ & $\oplus $ & $16\times \overline{16}$ &  &  & 
\end{tabular}
\label{dec}
\end{equation}%
Notice that products like $10\times \overline{10},$ $10\times \overline{16}$
and so on can be also reduced to sums of irreducible representation of
SO(10). \newline
Thinking of the diagonal sub-blocks $L_{{\small 0}}^{{\small 0}}=\mathcal{N}%
_{\mathbf{1\times }\overline{\mathbf{1}}}$, $L_{{\small j}}^{{\small k}}=%
\mathcal{N}_{\mathbf{10\times }\overline{\mathbf{10}}}$ and $L_{{\small %
\beta }}^{{\small \gamma }}=\mathcal{N}_{\mathbf{16\times }\overline{\mathbf{%
16}}}$ in eq(\ref{ml}) as $R\otimes \bar{R}$ building blocks of SO$(2)\times 
$SO$\left( 10\right) $, we can represent the operator $\mathcal{L}_{E_{6}}$
in term\textrm{s} of the topological quiver Q$_{E_{6}}$ of the Figure 
\textbf{\ref{so10}}. It has three nodes $N_{i}$ and 3+3 oriented links 
\textsc{l}$_{ij}$ as depicted by the figure. The topological states
propagating in the quiver Q$_{E_{6}}$ are \emph{massless} as required by the
E$_{6}$ gauge symmetry of the topological 4D CS theory which has SO$%
(2)\times $SO$\left( 10\right) $ as a subsymmetry. Notice as well that all
states in the topological quiver of the Figure \textbf{\ref{so10}} are in
bi-representations $R_{i}\otimes \bar{R}_{j}$ of SO$(2)\times $SO$\left(
10\right) $. The three nodes N$_{i}$ are in self-dual representations in the
sense that $R_{i}$ and its transpose $\bar{R}_{j}$ have \textrm{the} same
dimension and are related by representation- duality. So, \textrm{the} nodes
can be imagined as describing self-dual topological matter of SO$(2)\times $%
SO$\left( 10\right) $. However, the links \textsc{l}$_{ij}$, though also
massless, are in oriented bi- representations $R_{i}\otimes \bar{R}_{j}$
with $R_{i}$ and $\bar{R}_{j}$ having different dimensions. In the Figure 
\textbf{\ref{so10}}, the 6 oriented links \textsc{l}$_{ij}$ and \textsc{l}$%
_{ji}$ between the three nodes are as follows: $\left( \mathbf{a}\right) $
the topological bi-matter $\mathcal{B}_{\mathbf{1\times }\overline{\mathbf{10%
}}}$ and $\mathcal{B}_{\mathbf{10\times }\overline{\mathbf{1}}}$
corresponding to $R_{1}$ and $R_{2}$ are given by the SO$(2)\times $SO$%
\left( 10\right) $ representations $\mathbf{1}_{-4/3}$ and $\mathbf{10}%
_{+2/3}$. $\left( \mathbf{b}\right) $ the topological bi-matter $\mathcal{B}%
_{\mathbf{1\times }\overline{\mathbf{16}}}$ and $\mathcal{B}_{\mathbf{%
16\times }\overline{\mathbf{1}}}$ with $R_{1}$ and $R_{2}$ \textrm{are}
given by the SO$\left( 10\right) $ representations $\mathbf{1}_{-4/3}$ and $%
\mathbf{16}_{-1/3}$. $\left( \mathbf{c}\right) $ the bi-matter $\mathcal{B}_{%
\mathbf{10\times }\overline{\mathbf{16}}}$ and $\mathcal{B}_{\mathbf{%
16\times }\overline{\mathbf{10}}}$ with $R_{1}$ and $R_{2}$ \textrm{are}
given by the representation $\mathbf{10}_{+2/3}$ and $\mathbf{16}_{-1/3}$. 
\newline
From (\ref{en}), we also learn the \textrm{two }following features
descending from the decomposition (\ref{dec}): $\left( \mathbf{i}\right) $
the bi-matters $\mathcal{B}_{\mathbf{1\times }\overline{\mathbf{16}}}$ and $%
\mathcal{B}_{\mathbf{16\times }\overline{\mathbf{1}}}$ between the nodes $%
\mathcal{N}_{\mathbf{1\times }\overline{\mathbf{1}}}$ and $\mathcal{N}_{%
\mathbf{16\times }\overline{\mathbf{16}}}$ can be interpreted as fundamental
(spinorial like) topological matter. \textrm{They are expressed in terms of}
Darboux coordinates $c_{\gamma }$ and $b^{\beta }$ \textrm{that} satisfy the
non trivial Poisson bracket 
\begin{equation}
\left \{ z^{-\frac{4}{3}}b^{\beta },z^{-\frac{4}{3}}c_{\gamma }\right \}
=z^{-\frac{8}{3}}\delta _{\gamma }^{\beta }  \label{bc}
\end{equation}%
$\left( \mathbf{ii}\right) $ the bi-matters $\mathcal{B}_{\mathbf{1\times }%
\overline{\mathbf{10}}}$ and $\mathcal{B}_{\mathbf{10\times }\overline{%
\mathbf{1}}}$ \textrm{linking} the nodes $\mathcal{N}_{\mathbf{1\times }%
\overline{\mathbf{1}}}$ and $\mathcal{N}_{\mathbf{10\times }\overline{%
\mathbf{10}}}$ can also be interpreted as fundamental (but vector like)
topological matter. They are given by two SO$\left( 10\right) $ vectors,
namely $V^{i}=\frac{1}{2}b^{\alpha }\left( \Gamma ^{i}\right) _{\alpha \beta
}b^{\beta }$ and $W_{i}=\frac{1}{2}c_{\alpha }\left( \Gamma _{i}\right)
^{\alpha \beta }c_{\beta }.$ These vectors are quadratic in the Darboux
coordinates \textrm{and} obey the non trivial Poisson bracket%
\begin{equation}
\left \{ z^{-\frac{4}{3}}V^{i},z^{-\frac{4}{3}}W_{j}\right \} =\frac{1}{2}%
\delta _{j}^{i}z^{-\frac{8}{3}}T+z^{-\frac{8}{3}}Z_{j}^{i}  \label{vw}
\end{equation}%
\textrm{where} T and Z$_{j}^{i}$ are quadratic in the Darboux coordinates,
they are given by the SO$\left( 2\right) \times $SO$\left( 10\right) $
scalar $T=b^{\alpha }c_{\alpha }$ and \textrm{the operator} $%
Z_{j}^{i}=b^{\beta }\Omega _{j\beta }^{i\alpha }c_{\alpha }$ with 
\begin{equation}
\Omega _{j\beta }^{i\alpha }=\frac{1}{2}(\tilde{\Gamma}^{i}\Gamma
_{j}+\Gamma _{j}\tilde{\Gamma}^{i})_{\beta }^{\alpha }
\end{equation}%
\textrm{and} $\tilde{\Gamma}^{i}$ referring to the transpose of the Gamma
matrix $\Gamma ^{i}.$ The underlying properties of the Poisson brackets (\ref%
{bc}-\ref{vw}) and other aspects will be reported in future occasion.

\section{Minuscule CGY operator $\mathcal{L}_{E_{7}}$}

In this section, we construct the 't Hooft line operator $\mathcal{L}%
_{E_{7}} $ of \textrm{the} 4D CS theory with E$_{7}$ gauge symmetry \textrm{%
and} underlying \textrm{Lie algebra }$e_{7}$. We also give the topological
gauge quiver Q$_{E_{7}}$ associated with the $\mathcal{L}_{E_{7}}$ \textrm{%
by following} the approach used above for the topological E$_{6}$ theory.

\subsection{Levi subalgebra and weights of the \textbf{56}$_{\mathbf{e}_{7}}$%
}

\begin{figure}[tbph]
\begin{center}
\includegraphics[width=8cm]{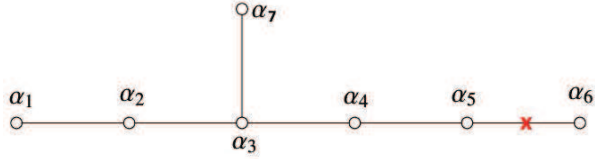}
\end{center}
\par
\vspace{-0.5cm}
\caption{Dynkin Diagram of E$_{7}$ having seven nodes labeled by the simple
roots $\protect \alpha _{i}$. The cross $\left( \times \right) $ indicates
the roots used in the Levi decomposition with Levi subgroup $SO(2)\times
E_{6}.$}
\label{E7}
\end{figure}

\begin{figure}[tbph]
\begin{center}
\includegraphics[width=12cm]{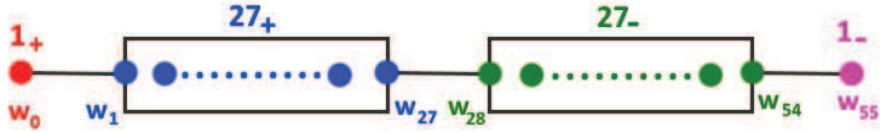}
\end{center}
\par
\vspace{-0.5cm}
\caption{The decomposition of the $\mathbf{56}$ representation of E$_{7}$ in
terms of representations of E$_{6}$. We have $\mathbf{56}=\mathbf{28}%
_{+q}\oplus \mathbf{28}_{-q}$ with $\mathbf{28}_{\pm q}$ reducible like $%
\mathbf{1}_{\pm 3/2}\oplus \mathbf{27}_{\pm 1/2}.$}
\label{127}
\end{figure}

We begin by giving some useful ingredients concerning the exceptional Lie
algebra $\boldsymbol{e}_{7}$. First, the root system $\Phi _{\boldsymbol{e}%
_{7}}$ of this Lie algebra contains 126 roots; half of them \textrm{are}
positive and the other\textrm{s are} negative. This system is a subset of
the $\Phi _{\boldsymbol{e}_{8}}$ system of the exceptional Lie algebra $%
\boldsymbol{e}_{8}$ containing 240 roots; 126 of them sit in its subset $%
\Phi _{\boldsymbol{e}_{7}}$ and the \textrm{rest} in its complement $\Phi _{%
\boldsymbol{e}_{8}}\backslash \Phi _{\boldsymbol{e}_{7}}$. This feature
implies that $\Phi _{\boldsymbol{e}_{7}}$ can be derived from $\Phi _{%
\boldsymbol{e}_{8}}$ which is built by using an eight dimensional vector
basis $\left \{ \epsilon _{1},...,\epsilon _{8}\right \} $ of $\mathbb{R}%
^{8} $ with $\epsilon _{i}.\epsilon _{j}=\delta _{ij}$. Within this view,
the seven simple roots can be taken as follows: the \textrm{first simple
root is equivalent to the one in} (\ref{al1})%
\begin{equation}
E_{7}:\alpha _{1}=\frac{1}{2}\left( \epsilon _{1}-\epsilon _{2}-\epsilon
_{3}-\epsilon _{4}-\epsilon _{5}-\epsilon _{6}-\epsilon _{7}+\epsilon
_{8}\right)  \label{a}
\end{equation}%
with five roots defined as $\alpha _{i}=\epsilon _{i}-\epsilon _{i-1}$ for $%
2\leq i\leq 6$ and the seventh one as $\alpha _{7}=\epsilon _{1}+\epsilon
_{2}.$ The Dynkin diagram associated with these roots is given by the Figure 
\textbf{\ref{E7},} from which we \textrm{can} determine the Cartan matrix $%
K_{e_{7}}.$ 
The 72 roots of $\boldsymbol{e}_{6}$ split into 36 positive roots and 36
negative ones. The 36 positive roots split in turn as 20+16. The 20 positive
roots are given by $\epsilon _{i}\pm \epsilon _{j}$ with $1\leq j<i\leq 5$
and the 16 ones read as $\frac{1}{2}q_{i}\epsilon _{i}+\epsilon
_{6}-\epsilon _{7}+\epsilon _{8}$ with the five $q_{i}=\pm 1$ constrained as 
$\Pi _{i}q_{i}=-1.$\newline
As far as the root system of $\boldsymbol{e}_{7}$ is concerned, notice that
contrary to the six simple roots $\alpha _{2},...,\alpha _{7}$ constructed
out of the six $\epsilon _{1},...,\epsilon _{6}$, the simple $\alpha _{1}$
has a spinorial-like nature, it moreover depends on two extra dimensions
generated by $\epsilon _{7}$ and $\epsilon _{8}.$ This extra dependence is
carried by the particular combination $\beta _{\max }=\epsilon _{8}-\epsilon
_{7}$ that is \textrm{expressed} like%
\begin{equation}
\beta _{\max }=2\alpha _{1}+3\alpha _{2}+4\alpha _{3}+3\alpha _{4}+2\alpha
_{5}+\alpha _{6}+2\alpha _{7}  \label{g}
\end{equation}%
with length $\beta _{\max }^{2}=2.$ \textrm{This is the highest positive
root, and it plays an important role} in the breaking $E_{6}\rightarrow
SO\left( 2\right) \times SO\left( 10\right) .$ Regarding the
Levi-decomposition $\boldsymbol{e}_{7}=\boldsymbol{n}_{-}\oplus \boldsymbol{l%
}_{\mu }\oplus \boldsymbol{n}_{+}$, recall that the exceptional $\boldsymbol{%
e}_{7}$ has one minuscule coweight $\mu $ given by $\lambda _{6}=\epsilon
_{6}+\beta _{\max }/2$ obeying $\lambda _{6}.\alpha _{i}=\delta _{i6}.$ It
corresponds to the fundamental representation \textbf{56} of $\boldsymbol{e}%
_{7}$ which is \textrm{self dual and pseudo-real} \textrm{\cite{sld}}.
Recall also that the Levi- subalgebra is given by $\boldsymbol{l}_{\mu
}=so(2)\oplus e_{6}$ and the nilpotents are $\boldsymbol{n}_{+}=27_{+}$ and $%
\boldsymbol{n}_{-}=27_{-}.$ So, the Levi- decomposition of the Lie algebra $%
\boldsymbol{e}_{7}$ dispatches its 133 dimensions in terms of
representations of $\boldsymbol{e}_{6}$ like $1+78+27_{+}+27_{-}$.
Similarly, the fundamental \textbf{56} representation of $\boldsymbol{e}%
_{7}, $ characterising the 't Hooft line operator of the topological E$_{7}$
theory, decomposes with respect to $so(2)\oplus e_{6}$ as follows \textrm{%
\cite{br}}%
\begin{equation}
\mathbf{56}=\mathbf{28}_{+q}\oplus \mathbf{28}_{-q}\qquad ,\qquad \mathbf{28}%
_{\pm q}=\mathbf{1}_{\pm q}\oplus \mathbf{27}_{\pm q}  \label{28}
\end{equation}%
This representation of E$_{7}$ is made of four E$_{6}$- representations: 
\textrm{two singlets} $\mathbf{1}_{\pm 3/2}$ with so(2) charges $\pm 3/2$;
and two fundamentals $\mathbf{27}_{\pm 1/2}$ with so(2) charges $\pm 1/2$.
To deal with the 56 states $\left \{ \left \vert \omega _{i}\right \rangle
\right \} _{0\leq i\leq 55}$ of this fundamental representation, we use the
diagram of the Figure \textbf{\ref{127}} to label the 28 weights of $\mathbf{%
28}_{+q}$ by the subset $W_{+}=\left \{ \left \vert \omega
_{i}\right
\rangle \right \} _{0\leq i\leq 27}$; and the 28 weights of the $%
\mathbf{28}_{-q}$ by $W_{-}=\left \{ \left \vert \omega _{i}\right \rangle
\right \} _{28\leq i\leq 55}$. 
Generic weights $\omega _{i}$ in W$_{+}\cup $W$_{-}$ obey some special
features that are useful for the construction of the operator $\mathcal{L}%
_{E_{7}}$, we give some of them \textrm{here.} First, we have $\omega
_{27}=\omega _{0}-\beta _{\max }$ and $\omega _{28}=\omega _{55}+\beta
_{\max }$, \textrm{from which }we learn the interesting relation $\omega
_{27}+\omega _{28}=\omega _{0}+\omega _{55}$ relating "boundary" weights in W%
$_{\pm }.$ This feature is in fact a general property of weights $\omega
_{i} $ in \textbf{56}, it extends like $\omega _{i}+\omega _{55-i}=\omega
_{0}+\omega _{55}$ for the label $i$ ranging the full interval $0\leq i\leq
27$. We also have the \textrm{relations} $\omega _{i}=\omega _{0}-\gamma
_{i} $ and $\omega _{55-i}=\omega _{55}+\gamma _{i}$ for generic roots $%
\gamma _{i}$ in the nilpotent $\mathbf{27}_{+}$. In these regards, recall
that the highest weight state of the representation \textbf{56} is given by $%
\left \vert \omega _{0}\right \rangle =\left \vert \lambda
_{6}\right
\rangle $ and its lowest weight state is given by $\left \vert
\omega _{55}\right \rangle =\left \vert -\lambda _{6}\right \rangle .$ As
such, we have the following \textrm{additional} properties: $\left( \mathbf{a%
}\right) $ The lowest and the highest boundary weights obey $\omega
_{55}+\omega _{0}=0$ and $\omega _{0}-\omega _{55}=2\lambda _{6}.$ $\left( 
\mathbf{b}\right) $ For generic weights, we have the relations $\omega
_{55-i}=\omega _{55}+\gamma _{i}$ and so $\omega _{i}-\omega
_{55-i}=2\lambda _{6}-2\gamma _{i}$ for $0\leq i\leq 27$ with $\gamma
_{0}\equiv 0.$ For convenience, we use the following simple notations: 
\newline
$\left( {\small 1}\right) $ the weights $\left \vert \omega
_{l}\right
\rangle $ in W$_{+}$ are denoted like $\left \vert
l_{+}\right
\rangle $ with $0_{+}\leq l_{+}\leq 27_{+}.$ \newline
$\left( {\small 2}\right) $ the weights $\left \vert \omega
_{l}\right
\rangle $ in W$_{-}$ with $28\leq l\leq 55$ are denoted like $%
\left \vert l_{-}\right \rangle $ with $27_{-}\leq l_{-}\leq 0_{-}$ where $%
\left \vert \omega _{55}\right \rangle =\left \vert 0_{-}\right \rangle .$ 
\newline
$\left( {\small 3}\right) $ For a positive root $\beta _{s}$ belonging to $%
\Phi _{e_{7}}\backslash \Phi _{e_{6}},$ the states $X_{\beta
_{s}}\left
\vert \omega _{0}\right \rangle =\left \vert \omega _{0}-\beta
_{s}\right
\rangle $ and $Y^{\beta _{s}}\left \vert \omega
_{55}\right
\rangle =\left
\vert \omega _{55}+\beta _{s}\right \rangle $
are denoted like $X_{s}\left
\vert 0\right \rangle =\left \vert
s_{+}\right
\rangle $ and $Y^{s}\left
\vert 0_{-}\right \rangle
=\left
\vert s_{-}\right \rangle $ with integer $0\leq s\leq 27$.

\subsection{Constructing the $\mathcal{L}_{E_{7}}$- operator and its Q$%
_{E_{7}}$}

The construction of the operator $\mathcal{L}_{E_{7}}$ follows the same
method we have used for the derivation of $\mathcal{L}_{E_{6}}$. However,
contrary to the E$_{6}$ topological theory, the $\mathcal{L}_{E_{7}}$ of the
4D Chern-Simons theory with E$_{7}$ gauge symmetry has \textrm{somehow
specific features}. This is because of the self- duality and the
pseudo-reality of the fundamental\textbf{\ 56} representation of E$_{7}$
encoded in the splitting $\left( 28_{+},28_{-}\right) $. These particular
properties lead to a remarkable expression of the operator\textrm{\ }$%
\mathcal{L}_{E_{7}}$. Below, we describe the main lines for the derivation
of the explicit value of $\mathcal{L}_{E_{7}}$ and the associated
topological gauge quiver Q$_{E_{7}}$. \textrm{Further details are reported
in the appendix}. To fix the ideas, we think it is interesting to anticipate
the study of $\mathcal{L}_{E_{7}}$ by giving first the structure of the
obtained quiver Q$_{E_{7}}$ and turn later to comment it. 
\begin{figure}[tbph]
\begin{center}
\includegraphics[width=5cm]{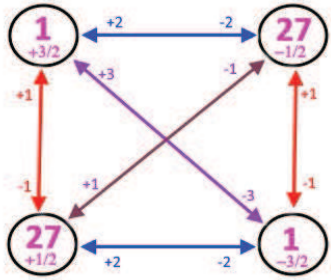}
\end{center}
\par
\vspace{-0.5cm}
\caption{The topological quiver Q$_{E_{7}}$ representing $\mathcal{L}%
_{E_{7}} $. It has 4 nodes and 12 links. The nodes describe self-dual
topological matter uncharged under $SO\left( 2\right) $. The links describe
bi-matter in $\left( R_{i},\bar{R}_{j}\right) $ of $E_{6}$ charged under SO$%
\left( 2\right) $ with charges $\pm 1,\pm 2,$ $\pm 3.$}
\label{2828}
\end{figure}
It has four nodes describing and 6+6 oriented links as depicted by the Figure%
\textrm{\ }\textbf{\ref{2828}.} It has an outer-automorphism symmetry
exchanging the nodes $\mathbf{1}_{{\small +3/2}}$ and $\mathbf{1}_{{\small %
-3/2}}$ as well as the $\mathbf{27}_{{\small +1/2}}$ and $\mathbf{27}_{%
{\small -1/2}}$. Recall that the $\mathcal{L}_{E_{7}}$ we want to build is
given by $e^{X}z^{\mathbf{\mu }}e^{Y}$ with diagonal $\mathbf{\mu }$ sitting
in the Levi- subalgebra $so(2)\oplus e_{6}.$ The $X=b^{\beta }X_{\beta }$ is
valued in the nilpotent $\boldsymbol{27}_{+}$ and the $Y=c_{\beta }Y^{\beta
} $ in the $\boldsymbol{27}_{-}$ with positive root $\beta $ belonging to
the root subsystem $\Phi _{\boldsymbol{e}_{7}}\backslash \Phi _{\boldsymbol{e%
}_{6}}.$ The generators obey the commutation relations $\left[ \mathbf{\mu ,}%
X_{\beta }\right] =X_{\beta }$ and $\left[ \mathbf{\mu ,}Y^{\beta }\right]
=-Y^{\beta }$ from which we deduce 
\begin{equation}
\left[ \mathbf{\mu ,}X\right] =X\qquad ,\qquad \left[ \mathbf{\mu ,}Y\right]
=-Y  \label{com}
\end{equation}%
So, the $\mathbf{\mu }$ is a diagonal charge operator; it can be decomposed
like $\mathbf{\mu }=\mu _{\mathbf{1}_{+}}+\mu _{\mathbf{27}_{+}}+\mu _{%
\mathbf{27}_{-}}+\mu _{\mathbf{1}_{-}}$ with $\mu _{\mathbf{1}_{q}}=\mu _{%
\mathbf{1}_{q}}^{0_{q}}\varrho _{0_{q}}$ where $\varrho _{0_{q}}=\left \vert
\omega _{0_{q}}\right \rangle \left \langle \omega _{0_{q}}\right \vert $
and $\mu _{\mathbf{1}_{q}}^{0_{q}}=3q/2$. Similar expressions can be written
down for $\mu _{\mathbf{27}_{q}}$ \textrm{where} we have $\mu _{\mathbf{27}%
_{q}}=\mu _{\mathbf{27}_{q}}^{\alpha _{q}}\varrho _{\alpha _{q}}$ such that $%
\varrho _{\alpha _{q}}=\left \vert \omega _{\alpha _{q}}\right \rangle
\left
\langle \omega _{\alpha _{q}}\right \vert $ and $\mu _{\mathbf{27}%
_{q}}^{\alpha _{q}}=q/2$ with $q=\pm 1$. Formally, we have \textrm{\cite{bra}%
} 
\begin{equation}
\mathbf{\mu }=\frac{3}{2}\varrho _{\mathbf{1}_{+}}+\frac{1}{2}\varrho _{%
\mathbf{27}_{+}}-\frac{1}{2}\varrho _{\mathbf{27}_{-}}-\frac{3}{2}\varrho _{%
\mathbf{1}_{-}}  \label{72}
\end{equation}%
where the charges $\pm 3/2$ and $\pm 1/2$ are \textrm{associated to} the
four nodes of the quiver Q$_{E_{7}}.$ For a given positive root $\beta $ in $%
\Phi _{e_{7}}\backslash \Phi _{e_{6}}$, the generators $X_{\beta }$ and $%
Y^{\beta }$ solving (\ref{com}) are as follows 
\begin{equation}
\begin{tabular}{lll}
$X_{\beta }$ & $=$ & $\left \vert \omega _{0_{+}}\right \rangle \left
\langle \omega _{\beta _{+}}\right \vert +\left \vert \omega _{\delta
_{+}}\right \rangle \Gamma _{\beta }^{\delta _{+}\gamma _{-}}\left \langle
\omega _{\gamma _{-}}\right \vert +\left \vert \omega _{\beta _{-}}\right
\rangle \left \langle \omega _{0_{-}}\right \vert $ \\ 
$Y^{\beta }$ & $=$ & $\left \vert \omega _{0_{-}}\right \rangle \left
\langle \omega ^{\beta _{-}}\right \vert +\left \vert \omega ^{\gamma
_{-}}\right \rangle \bar{\Gamma}_{\gamma _{-}\delta _{+}}^{\beta }\left
\langle \omega ^{\delta _{+}}\right \vert +\left \vert \omega ^{\beta
_{+}}\right \rangle \left \langle \omega _{0_{+}}\right \vert $%
\end{tabular}
\label{xay}
\end{equation}%
where $\Gamma _{\beta }^{\delta _{+}\gamma _{-}}$ and $\bar{\Gamma}_{\gamma
_{-}\delta _{+}}^{\beta }$ are coupling tensors of three \textbf{27}
representations of E$_{6}$. They are respectively given by $\left \langle
\omega ^{\delta _{+}}|X_{\beta }|\omega ^{\gamma _{-}}\right \rangle $ and $%
\left \langle \omega _{\gamma _{-}}|Y^{\beta }|\omega _{\delta
_{+}}\right
\rangle $. By using these expressions, we determine the
corresponding Cartan charge operator $\frac{2}{3}H_{\beta }=\left[ X_{\beta
},Y^{\beta }\right] $ from which we deduce the relation $\Gamma _{\beta
}^{\delta _{+}\gamma _{-}}\bar{\Gamma}_{\gamma _{-}\eta _{+}}^{\beta }=\frac{%
4}{3}\delta _{\beta }^{\delta _{+}}\delta _{\eta _{+}}^{\beta }.$
Multiplying (\ref{xay}) respectively by Darboux coordinates $b^{\beta }$ and 
$c_{\beta },$ we deduce the expressions of the matrix operators $X=b^{\beta
}X_{\beta }$ and $Y=c_{\beta }Y^{\beta }$ needed for the calculation of $%
e^{X}z^{\mathbf{\mu }}e^{Y}.$ Using these expressions, we compute the powers
X$^{n}$ and Y$^{n}$; we find that $X^{4}=Y^{4}=0$ and (\textrm{see Appendix
for more details}),%
\begin{equation}
\begin{tabular}{lllllll}
$X^{2}$ & $=$ & $2S^{\beta _{-}}\left \vert \omega _{0_{+}}\right \rangle
\left \langle \omega _{\beta _{-}}\right \vert +2S^{\beta _{+}}\left \vert
\omega _{\beta _{+}}\right \rangle \left \langle \omega _{0_{-}}\right \vert 
$ & , & $X^{3}$ & $=$ & $6\mathcal{E}\left \vert \omega _{0_{+}}\right
\rangle \left \langle \omega _{0_{-}}\right \vert $ \\ 
$Y^{2}$ & $=$ & $2R_{\alpha _{+}}\left \vert \omega _{0_{-}}\right \rangle
\left \langle \omega ^{\alpha _{+}}\right \vert +2R_{\alpha _{-}}\left \vert
\omega ^{\alpha _{-}}\right \rangle \left \langle \omega _{0_{+}}\right
\vert $ & , & $Y^{3}$ & $=$ & $6\mathcal{F}\left \vert \omega _{0_{-}}\right
\rangle \left \langle \omega _{0_{+}}\right \vert $%
\end{tabular}%
\end{equation}%
where for convenience we have set%
\begin{equation}
\begin{tabular}{lllllllllll}
$S^{\beta _{-}}$ & $=$ & $\frac{1}{2}b^{\gamma }\Gamma _{\gamma _{+}\delta
}^{\beta _{-}}b^{\delta }$ & , & $R_{\alpha _{-}}\ $ & $=$ & $\frac{1}{2}%
c_{\gamma }\Gamma _{\alpha _{-}}^{\gamma \delta _{+}}c_{\delta }$ & , & $%
\mathcal{E}$ & $=$ & $\frac{1}{3}b_{\beta _{+}}S^{\beta _{+}}$ \\ 
$S^{\beta _{+}}$ & $=$ & $\frac{1}{2}b^{\gamma }\Gamma _{\gamma \delta
_{-}}^{\beta _{+}}b^{\delta }$ & , & $R_{\alpha _{+}}$ & $=$ & $\frac{1}{2}%
c_{\gamma }\Gamma _{\alpha _{+}}^{\gamma _{-}\delta }c_{\delta }$ & , & $%
\mathcal{F}$ & $=$ & $\frac{1}{3}R_{\alpha _{-}}c^{\alpha _{-}}$%
\end{tabular}
\label{sr}
\end{equation}%
\textrm{Notice as well}, the following features regarding the quantities (%
\ref{sr}) which are involved in e$^{X}$ and e$^{Y}$ and which will appear in 
$\mathcal{L}_{E_{7}}$ and its quiver Q$_{E_{7}}$: $\left( \mathbf{i}\right) $
The $S^{\beta _{\pm }}$'s are quadratic in $b^{\gamma }$ and carry a charge $%
-2$ under SO(2). This is because the b$^{\alpha }$'s carry a charge $-1$ and
the c$_{\alpha }$'s carry a charge $+1$. $\left( \mathbf{ii}\right) $\ The $%
R_{\alpha _{\pm }}$'s are quadratic in $c_{\gamma }$ and carry a charge $+2$
under SO(2). $\left( \mathbf{iii}\right) $ The $\mathcal{E}$ and the $%
\mathcal{F}$ are respectively cubic in $b^{\gamma }$ and $c_{\gamma }$; they
carry charges $-3$ and $+3$ under SO(2). \textrm{In this regard, observe that%
} $\mathcal{E}$ and $\mathcal{F}$ are scalars of $\boldsymbol{e}_{6}$ and
can be put in the form $\mathcal{E}=\frac{1}{6}\Gamma _{\alpha \beta \gamma
}b^{\alpha }b^{\beta }b^{\gamma }$ and $\mathcal{F}=\frac{1}{6}\bar{\Gamma}%
^{\alpha \beta \gamma }c_{\alpha }c_{\beta }c_{\gamma }$ as shown by eqs(\ref%
{3X}-\ref{3Y}). Using these relations and setting $T_{\alpha _{+}}^{\beta
_{-}}=b^{\eta }\Gamma _{\eta \alpha _{+}}^{\beta _{-}}$ and $J_{\alpha
_{-}}^{\beta _{+}}=c_{\eta }\Gamma _{\alpha _{-}}^{\eta \beta _{+}},$ we can
determine the explicit expressions of the exponentials $e^{X}$ and $z^{\mu
}e^{Y}$ in the vector basis of the 56 weights ordered like $\left \{
\left
\vert \omega _{0_{+}}\right \rangle ,\left \vert \omega _{\beta
_{+}}\right
\rangle ,\left \vert \omega _{\beta _{-}}\right \rangle
\left
\vert \omega _{0_{-}}\right \rangle \right \} .$ We find that $e^{X}$
and $z^{\mu }e^{Y}$ are \textrm{respectively equal to}%
\begin{equation}
e^{X}=\left( 
\begin{array}{cccc}
1 & b^{\beta _{+}} & S^{\beta _{-}} & \mathcal{E} \\ 
0 & \delta _{\alpha _{+}}^{\beta _{+}} & T_{\alpha _{+}}^{\beta _{-}} & 
S_{\alpha _{+}} \\ 
0 & 0 & \delta _{\alpha _{-}}^{\beta _{-}} & b_{\alpha _{-}} \\ 
0 & 0 & 0 & 1%
\end{array}%
\right) ,\quad z^{\mu }e^{Y}=\left( 
\begin{array}{cccc}
z^{\frac{3}{2}} & 0 & 0 & 0 \\ 
z^{\frac{1}{2}}c_{\alpha _{+}} & z^{\frac{1}{2}}\delta _{\alpha _{+}}^{\beta
_{+}} & 0 & 0 \\ 
z^{-\frac{1}{2}}R_{\alpha _{-}} & z^{-\frac{1}{2}}J_{\alpha _{-}}^{\beta
_{+}} & z^{-\frac{1}{2}}\delta _{\alpha _{-}}^{\beta _{-}} & 0 \\ 
z^{-\frac{3}{2}}\mathcal{F} & z^{-\frac{3}{2}}R^{\beta _{+}} & z^{-\frac{3}{2%
}}c^{\beta _{-}} & z^{-\frac{3}{2}}%
\end{array}%
\right)
\end{equation}%
Putting these relations into $e^{X}z^{\mu }e^{Y}$, we obtain the explicit
expression of the 't Hooft line operator $\mathcal{L}_{E_{7}}$ that we
present as follows%
\begin{equation}
\mathcal{L}_{E_{7}}=\left( 
\begin{array}{cccc}
L_{0_{+}}^{0_{+}} & L_{0_{+}}^{\beta _{+}} & L_{0_{+}}^{\beta _{-}} & z^{-%
\frac{3}{2}}\mathcal{E} \\ 
L_{\alpha _{+}}^{0_{+}} & L_{\alpha _{+}}^{\beta _{+}} & L_{\alpha
_{+}}^{\beta _{-}} & z^{-\frac{3}{2}}S_{\alpha _{+}} \\ 
L_{\alpha _{-}}^{0_{+}} & L_{\alpha _{-}}^{\beta _{+}} & L_{\alpha
_{-}}^{\beta _{-}} & z^{-\frac{3}{2}}b_{\alpha _{-}} \\ 
z^{-\frac{3}{2}}\mathcal{F} & z^{-\frac{3}{2}}R^{\beta _{+}} & z^{-\frac{3}{2%
}}c^{\beta _{-}} & z^{-\frac{3}{2}}%
\end{array}%
\right)  \label{le7}
\end{equation}%
The four diagonal sub-blocks $L_{0_{+}}^{0_{+}},$ $L_{\alpha _{+}}^{\beta
_{+}},$ $L_{\alpha _{-}}^{\beta _{-}}$ and $L_{0_{-}}^{0_{-}}$ are given by%
\begin{equation}
\begin{tabular}{lll}
$L_{0_{+}}^{0_{+}}$ & $=$ & $z^{\frac{3}{2}}+z^{\frac{1}{2}}b^{\alpha
_{+}}c_{\alpha _{+}}+z^{-\frac{1}{2}}S^{\alpha _{-}}R_{\alpha _{-}}+z^{-%
\frac{3}{2}}\mathcal{EF}$ \\ 
$L_{\alpha _{+}}^{\beta _{+}}$ & $=$ & $z^{\frac{1}{2}}\delta _{\alpha
_{+}}^{\beta _{+}}+z^{-\frac{1}{2}}T_{\alpha _{+}}^{\gamma _{-}}J_{\gamma
_{-}}^{\beta _{+}}+z^{-\frac{3}{2}}S_{\alpha _{+}}R^{\beta _{+}}$ \\ 
$L_{\alpha _{-}}^{\beta _{-}}$ & $=$ & $z^{-\frac{1}{2}}\delta _{\alpha
_{-}}^{\beta _{-}}+z^{-\frac{3}{2}}b_{\alpha _{-}}c^{\beta _{-}}$%
\end{tabular}%
\end{equation}%
and the off diagonal terms by%
\begin{equation}
\begin{tabular}{lllllll}
${\small L}_{{\small 0}_{+}}^{{\small \beta }_{+}}$ & $=$ & ${\small z}^{%
\frac{1}{2}}{\small b}^{\beta _{+}}{\small +z}^{-\frac{1}{2}}{\small S}^{%
{\small \alpha }_{-}}{\small J}_{{\small \alpha }_{-}}^{{\small \beta }_{+}}%
{\small +z}^{-\frac{3}{2}}\mathcal{E}R^{{\small \beta }_{+}}$ & , & $%
L_{\alpha _{-}}^{0_{+}}$ & $=$ & $z^{-\frac{1}{2}}R_{\alpha _{-}}+z^{-\frac{3%
}{2}}b_{\alpha _{-}}\mathcal{F}$ \\ 
${\small L}_{{\small 0}_{+}}^{{\small \beta }_{-}}$ & $=$ & $z^{-\frac{1}{2}%
}S^{\beta _{-}}+z^{-\frac{3}{2}}\mathcal{E}c^{\beta _{-}}$ & , & $L_{\alpha
_{+}}^{\beta _{-}}$ & $=$ & $z^{-\frac{1}{2}}T_{\alpha _{+}}^{\beta
_{-}}+z^{-\frac{3}{2}}S_{\alpha _{+}}c^{\beta _{-}}$ \\ 
${\small L}_{{\small \alpha }_{+}}^{{\small 0}_{+}}$ & $=$ & $z^{\frac{1}{2}%
}c_{{\small \alpha }_{+}}+z^{-\frac{1}{2}}T_{{\small \alpha }_{+}}^{{\small %
\gamma }_{-}}R_{{\small \gamma }_{-}}+z^{-\frac{3}{2}}S_{{\small \alpha }%
_{+}}\mathcal{F}$ & , & $L_{\alpha _{-}}^{\beta _{+}}$ & $=$ & $z^{-\frac{1}{%
2}}J_{\alpha _{-}}^{\beta _{+}}+z^{-\frac{3}{2}}b_{\alpha _{-}}R^{\beta
_{+}} $%
\end{tabular}
\label{2e7}
\end{equation}%
\textrm{For more details concerning these calculations, see the appendix.}
We end this description by noticing that the topological gauge quiver Q$%
_{E_{7}}$ representing the 't Hooft line operator (\ref{le7}) is given by
the Figure \textbf{\ref{2828}}. The construction of this topological quiver
is obtained by following the same method we have used for the building of Q$%
_{E_{6}}$ of the Figure \textbf{\ref{so10}}. For the case of Q$_{E_{7}},$
the Darboux coordinates $\left( b^{\alpha },c_{\alpha }\right) $ also have
an interpretation in terms of fundamental matter carrying a unit charge
under SO(2). We also find that $S_{\alpha }$ and $R^{\alpha }$ describe
fundamental matter; but with SO$\left( 2\right) $ charges $q=\pm 2;$ see the
blue links in the quiver Q$_{E_{7}}$ given by the Figure \textbf{\ref{2828}}.

\section{Conclusion}

Two- dimensional integrable field theories and integrable spin models
represent a significant area in classical and quantum physics including 2D
critical phenomena. They still bear\ several open questions intending to
explicitly describe the interactions between fundamental particles and
topological lines. In this paper, we have contributed in this matter by
considering topological 4D- CS theory in presence of exceptional minuscule
't Hooft lines. This particular topological theory has gauge symmetries
given by the E$_{6}$ and E$_{7}$ groups with 't Hooft lines described by the
complex representation \textbf{27} of E$_{6}$ and the self-dual \textbf{56}
representation of E$_{7}$. To undertake this study, we first revisited
useful aspects on the $\mathcal{L}_{G}$- operators in topological 4D CS
theory with gauge symmetry G by following the approach of \textrm{\cite%
{1F,1A,2A}}. Then, we focused on the E$_{6}$ and E$_{7}$ theories and
derived the explicit oscillator realisation of the corresponding $\mathcal{L}%
_{E_{6}}$- and $\mathcal{L}_{E_{7}}$- operators where specific properties
for exceptional groups have been found. The order $\eta $ of nilpotency of
the X$^{\eta }$ and Y$^{\eta }$ matrices in (\ref{xy}) is equal to 3 for E$%
_{6}$ versus $\eta =2$ for A- theory. It is equal to 4 for the case of E$%
_{7} $- theory. The oscillator realisation of the $\mathcal{L}$- operator of
the E$_{6}$ gauge symmetry is given by eq(\ref{en}), and the representation
of the $\mathcal{L}_{E_{7}}$ is given by eqs(\ref{le7}-\ref{2e7}). We also
proposed a graph to represent the operators $\mathcal{L}_{E_{6}}$ and $%
\mathcal{L}_{E_{7}}$ using topological gauge quivers Q$_{E_{6}}$ and Q$%
_{E_{7}}$ given by the Figures \textbf{\ref{so10}} and \textbf{\ref{2828}}.
In this diagrammatic representation, the Darboux coordinates have been
interpreted as topological fundamental matter and the nodes as topological
self-dual matter. General aspects of the topological gauge quivers Q$_{G}$
in 4D- CS theory with generic gauge symmetry G and the underlying algebra of
its nodes and links will be reported in future occasion. \textrm{It will be
also interesting to derive the Lax operators for the B- and C- types spin
chains obtained in \cite{Z} from 4D CS theory. An explicit investigation
regarding the derivation of Lax operators of A- , D- and E-types from 4D
Chern-Simons theory using topological quivers is given in \cite{Z2}. }

\section{Appendix}

In this appendix, we give details regarding the calculation of the Lax
operator (\ref{le7}-\ref{2e7}) derived from the 4D Chern-Simons theory with
gauge symmetry E$_{7}$. This operator $\mathcal{L}_{E_{7}}$ is given by the
formula $e^{X}z^{\mu }e^{Y}$ with $X$ and $Y$ valued in the nilpotent
subalgebras $\boldsymbol{n}_{+}=27_{+}$ and $\boldsymbol{n}_{-}=27_{-}$ of
the Levi- decomposition of the $e_{7}$ Lie algebra underlying the E$_{7}$
theory. Recall that in this decomposition, the Levi-subalgebra reads as $%
\boldsymbol{l}_{\mu }=so(2)\oplus e_{6}$ and the associated nilpotents are
given by $\boldsymbol{n}_{\pm }=27_{\pm }.$ To determine the explicit
expression of $\mathcal{L}_{E_{7}}$, we have to calculate $e^{X}z^{\mathbf{%
\mu }}e^{Y}$. This is done in three steps: $\left( i\right) $ We have to
work out the expression of the adjoint action of the minuscule coweight $%
\mathbf{\mu }$, it is given by (\ref{72}) which is equal to $\frac{3}{2}%
\varrho _{\mathbf{1}_{+}}+\frac{1}{2}\varrho _{\mathbf{27}_{+}}-\frac{1}{2}%
\varrho _{\mathbf{27}_{-}}-\frac{3}{2}\varrho _{\mathbf{1}_{-}}$. The
subscripts $\mathbf{R}$ appearing in the projectors $\varrho _{\mathbf{R}}$
refer to the decomposition (\ref{28}) of the fundamental representation of $%
e_{7}$ with respect to the representations of $so(2)\oplus e_{6},$ namely%
\begin{equation}
\mathbf{56}=\mathbf{1}_{3/2}\oplus \mathbf{27}_{+1/2}\oplus \mathbf{27}%
_{-1/2}\oplus \mathbf{1}_{-3/2}  \label{56}
\end{equation}%
$\left( \mathbf{ii}\right) $ We have to determine the expression of X and Y
solving the Levi-condition (\ref{com}) and then calculate the exponentials $%
e^{X}$ and $e^{Y}$ using the expansion $e^{Z}=\sum Z^{n}/n!.$ $\left( 
\mathbf{iii}\right) $ Once $\mathbf{\mu },$ $e^{X}$ and $e^{Y}$ are known,
we substitute into $e^{X}z^{\mu }e^{Y}$ and look for the explicit expression
of $\mathcal{L}_{E_{7}}$. Because of the special properties of the E$_{7}$
symmetry and its fundamental representation \textbf{56}, these calculations
are somehow technical in the sense that we need to exhibit features of the
root system $\Phi _{e_{7}}$ of $e_{7}$ and the weight vectors of its
fundamental \textbf{56}. Those useful tools for the calculation of $\mathcal{%
L}_{E_{7}}$ were reported in the main text; see subsection 5.1. Nevertheless
we think it interesting to give extra details lightening the explicit
computations like aspects concerning the Levi- decomposition and related
things for their role in the determination of $\mathcal{L}_{E_{7}}$. The $%
133 $ generators of $e_{7},$ its seven diagonal charge operators (rank) and
its 126 roots split under Levi-decomposition as collected in the following
table 
\begin{equation}
\begin{tabular}{|l|l|l|l|l|l|}
\hline
{\small algebra} & $\  \ e_{7}$ & $so_{2}$ & $\  \ e_{6}$ & $\  \  \boldsymbol{n}%
_{+}$ & $\  \  \boldsymbol{n}_{-}$ \\ \hline
{\small dim} & $\ {\small 133}$ & $\ {\small 1}$ & $\ {\small 78}$ & $\  \ 
{\small 27}$ & $\ {\small 27}$ \\ \hline
{\small rank} & $\  \ {\small 7}$ & $\ {\small 1}$ & $\  \ {\small 6}$ & $\  \
\ {\small 0}$ & $\  \ {\small 0}$ \\ \hline
{\small roots} & {\small 126} & \ {\small 0} & {\small 72} & \  \ {\small 27}
& \  \ {\small 27} \\ \hline
{\small Cartan H's} & $\text{{\small 7 H}}_{\text{{\small i}}}$ & {\small 1
H=}$\mathbf{\mu }$ & {\small 6}$\text{ }${\small H}$_{\text{{\small i}}}$ & 
\  \  \ 0 & \  \  \ 0 \\ \hline
{\small Step E's } & $\text{{\small 126 E}}_{{\small \pm }\alpha }$ & \ 0 & 
{\small 72}$\text{{\small \ E}}_{{\small \pm }\alpha }$ & {\small 27}$\text{%
{\small \ }}$X$_{{\small +}\beta }$ & {\small 27}$\text{{\small \ }}$Y$_{%
{\small -}\beta }$ \\ \hline
\end{tabular}
\label{t}
\end{equation}%
\begin{equation*}
\text{ \  \  \ }
\end{equation*}%
In this table, the n H$_{\text{{\small i}}}$ refers to the number n of
Cartan charge operators of the corresponding Lie algebra. Here the splitting
of the rank is given by $7=1+6$. The n\ E$^{{\small \pm }\alpha }$
represents the number n of step operators associated with the roots $\pm
\alpha $ of the Lie algebra. The splitting of the total roots is given by $%
126=72+(27+27)$. Regarding the technical details, notice the two following: $%
\left( \mathbf{1}\right) $ We have denoted the 27+27 generators of the
nilpotent algebras $\boldsymbol{n}_{+}$ and $\boldsymbol{n}_{-}$
respectively by $X_{\beta }$\textrm{\ and }$Y^{\beta }$ instead of the
conventional E$^{+\beta }$ and E$^{-\beta }$. For these generators, the
roots $\beta $ belong to the subset of positive roots $\Phi
_{e_{7}}^{+}\backslash \Phi _{e_{6}}^{+}$ introduced in the main text; that
is $\beta \in \Phi _{e_{7}}^{+}$ but $\beta \notin \Phi _{e_{6}}^{+};$ see
subsection 5.1. This discrimination in the notation is $\left( \mathbf{a}%
\right) $ because we have used it in the main text as given by the
expansions $X=b^{\beta }X_{\beta },$ $Y=c_{\beta }Y^{\beta }$ where $\beta $
is a positive root; and $\left( \mathbf{b}\right) $ in order to give an
interpretation of the $\Gamma _{\beta }^{\gamma _{+}\delta _{-}}$ used in
our calculations in terms of a 3-coupling of three $\boldsymbol{e}_{6}$
representations namely $27\times 27\times 27$. $\left( \mathbf{2}\right) $
The entries of the $X_{\beta }$ and $Y^{\beta }$ matrices namely $%
\left
\langle \omega _{\gamma _{+}}|X_{\beta }|\omega _{\delta
_{-}}\right
\rangle $\textrm{\ and }$\left \langle \omega _{\delta
_{-}}|Y^{\beta }|\omega _{_{+}\gamma }\right \rangle $\textrm{, with} kets $%
\left \vert \omega _{\eta }\right \rangle $ and bras $\left \langle \omega
_{\eta }\right \vert $ being 27 weight vector states in (\ref{56}) and their
27 duals, have been denoted in the text as $\Gamma _{\beta }^{\gamma
_{+}\delta _{-}}$ and $\bar{\Gamma}_{\delta _{-}\gamma _{+}}^{\beta }$. Here
also we have accommodated the labels of $\Gamma _{\beta }^{\gamma _{+}\delta
_{-}}$ and $\bar{\Gamma}_{\delta _{-}\gamma _{+}}^{\beta }$ in order to make
the result more accessible for the reader. The point is that the matrix
elements $\left
\langle \omega _{\gamma _{+}}|X_{\beta }|\omega _{\delta
_{-}}\right
\rangle $ and $\left \langle \omega _{\delta _{-}}|Y^{\beta
}|\omega _{\gamma _{+}}\right \rangle $ should respectively be written as $%
\left( X_{\beta }\right) _{\gamma _{+}}^{\delta _{-}}$ and $\left( Y^{\beta
}\right) _{\delta _{-}}^{\gamma _{+}};$ for convenience, we have used $%
\Gamma _{\beta }^{\gamma _{+}\delta _{-}}$ for $\left( X_{\beta }\right)
_{\gamma _{+}}^{\delta _{-}}$ and $\bar{\Gamma}_{\delta _{-}\gamma
_{+}}^{\beta }$ for $\left( Y^{\beta }\right) _{\delta _{-}}^{\gamma _{+}}.$
This technical detail is not very important as the calculations are
covariant. Notice that the label $\beta $ takes integer values from 1 to 27
as shown on the two last columns of the table (\ref{t}). Notice moreover
that the two other labels $\gamma _{+}$ and $\delta _{-}$ of the weights $%
\omega _{\gamma _{+}}$ and $\omega _{\delta _{-}}$ take also integer values
from 1 to 27 exactly like $\beta $. This is not a coincidence, the point is
that the set $\left
\{ \lambda \right \} _{e_{7}}$ of weights of the
adjoint representation adj$\boldsymbol{e}_{7}$ are precisely given by the
set $\Phi _{e_{7}}=\left \{ \alpha \right \} _{e_{7}}$ of roots of $%
\boldsymbol{e}_{7}$; that is $\left
\{ \lambda \right \} _{e_{7}}=\left \{
\alpha \right \} _{e_{7}}.$ Under the Levi-decomposition, we have, 
\begin{equation}
\left \{ \lambda \right \} _{adje_{7}}=\left \{ \lambda \right \}
_{adjso_{2}}\cup \left \{ \lambda \right \} _{adje_{6}}\cup \left \{ \lambda
\right \} _{\underline{27}}\cup \left \{ \lambda \right \} _{\overline{27}}
\end{equation}%
with the subset $\left \{ \lambda \right \} _{27}=\Phi
_{e_{7}}^{+}\backslash \Phi _{e_{6}}^{+}$ and the subset $\left \{ \lambda
\right \} _{\overline{27}}=\Phi _{e_{7}}^{-}\backslash \Phi _{e_{6}}^{-}.$
In other words, the 27 weights $\gamma _{+}=+\gamma $ belong to $\Phi
_{e_{7}}^{+}\backslash \Phi _{e_{6}}^{+}$ and the 27 weights $\delta
_{-}=-\delta $ sit in $\Phi _{e_{7}}^{-}\backslash \Phi _{e_{6}}^{-}.$ This
feature teaches us that $\gamma $ and $\delta $ belong also to $\Phi
_{e_{7}}^{+}\backslash \Phi _{e_{6}}^{+}$ exactly as $\beta $. \newline
After this general description concerning technicalities, we now turn to the
explicit computations of $\mathcal{L}_{E_{7}}$ by using the fundamental
coweight $\mathbf{\mu }$ given above. We start from the expansions $%
X=b^{\beta }X_{\beta }$ and $Y=c_{\beta }Y^{\beta }$ where $X_{\beta }$ and $%
Y^{\beta }$ are the generators of the nilpotent algebras in the Levi
decomposition $e_{7}\rightarrow so_{2}\oplus e_{6}\oplus 27_{+}\oplus
27_{-}. $ These generators solving the Levi- conditions are realised in the
weight basis $\left \{ \left \vert \omega _{0_{+}}\right \rangle
,\left
\vert \omega _{\delta _{+}}\right \rangle ,\left \vert \omega
_{\beta _{-}}\right \rangle ,\left \vert \omega _{0_{-}}\right \rangle
\right \} $ of the representation \textbf{56 }of\textbf{\ }$e_{7}$ as follows%
\begin{equation}
\begin{tabular}{lll}
$X_{\beta }$ & $=$ & $\left \vert \omega _{0_{+}}\right \rangle \left
\langle \omega _{\beta _{+}}\right \vert +\left \vert \omega _{\delta
_{+}}\right \rangle \Gamma _{\beta }^{\delta _{+}\gamma _{-}}\left \langle
\omega _{\gamma _{-}}\right \vert +\left \vert \omega _{\beta _{-}}\right
\rangle \left \langle \omega _{0_{-}}\right \vert $ \\ 
$Y^{\beta }$ & $=$ & $\left \vert \omega _{0_{-}}\right \rangle \left
\langle \omega ^{\beta _{-}}\right \vert +\left \vert \omega ^{\gamma
_{-}}\right \rangle \bar{\Gamma}_{\gamma _{-}\delta _{+}}^{\beta }\left
\langle \omega ^{\delta _{+}}\right \vert +\left \vert \omega ^{\beta
_{+}}\right \rangle \left \langle \omega _{0_{+}}\right \vert $%
\end{tabular}
\label{YX}
\end{equation}%
Here $\left \vert \omega _{0_{\pm }}\right \rangle $ refer to the singlets $%
\mathbf{1}_{\pm 3/2}$ in the decomposition (\ref{56}) and $\left \vert
\omega _{\delta _{\pm }}\right \rangle $ to the $\mathbf{27}_{\pm 1/2}.$ The
full list of the 56 weight vectors $\left \vert \omega
_{0_{+}}\right
\rangle ,\left \vert \omega _{\delta _{+}}\right \rangle
,\left \vert \omega _{\beta _{-}}\right \rangle ,\left \vert \omega
_{0_{-}}\right \rangle $ is known in the literature, it has been omitted
here for the simplicity of the presentation. However, we refer to \cite{sld}
for readers interested in this list. As described before, notice \textrm{%
that }$\Gamma _{\beta }^{\delta _{+}\gamma _{-}}$\textrm{\ in (\ref{YX}) is
given by }$\left \langle \omega _{\delta _{+}}|X_{\beta }|\omega _{\gamma
_{-}}\right \rangle $\textrm{\ and a} similar expression can be written for $%
\bar{\Gamma}_{\gamma _{-}\delta _{+}}^{\beta }.$ The tri-coupling $\Gamma
_{\beta }^{\delta _{+}\gamma _{-}}$ and $\bar{\Gamma}_{\gamma _{-}\delta
_{+}}^{\beta }$ are tensors with three labels $\left( \beta ,\gamma ,\delta
\right) $ all of them take integer values from 1 to 27. They have an
interpretation in terms of coupling three 27 (resp. $\overline{27}$ )
representations of e$_{6}$. In these regards, we recall that the
decomposition of the complex tensor product $27\times 27\times 27$ contains
the identity, and a similar property is valid for $\overline{27}\times 
\overline{27}\times \overline{27}$. Focussing on $27\times 27\times 27$,
this can be seen by first calculating the product $27\times 27=729$; it
decomposes like $729=\overline{27}+\overline{351}+\overline{351}^{\prime }.$
By putting into $27\times 27\times 27,$ we have $27\times \overline{27}%
+27\times \overline{351}+27\times \overline{351}^{\prime }$. Here, the
hermitian product $27\times \overline{27}$ decomposes like $1+78+650$
showing that the product of three 27s contains indeed the desired identity.%
\newline
Using the expressions given above, we can calculate the powers X$^{n}$ and Y$%
^{n}.$ We give below the calculations of X$^{n}.$ Similar computations are
valid for Y$^{n}$.

$\bullet $ \emph{Calculations of }$X^{2}$ \emph{and results for} $Y^{2}:$ By
substituting the expansion $X=b^{\beta }X_{\beta }$ into $X^{2},$ we have $%
X^{2}=b^{\alpha }b^{\beta }X_{\alpha }X_{\beta }$ with%
\begin{equation}
\begin{tabular}{lll}
$X_{\alpha }X_{\beta }$ & $=$ & $\left \vert \omega _{0_{+}}\right \rangle
\left \langle \omega _{\alpha _{+}}\right \vert \left( \left \vert \omega
_{0_{+}}\right \rangle \left \langle \omega _{\beta _{+}}\right \vert +\left
\vert \omega _{\delta _{+}}\right \rangle \Gamma _{\beta }^{\delta
_{+}\gamma _{-}}\left \langle \omega _{\gamma _{-}}\right \vert +\left \vert
\omega _{\beta _{-}}\right \rangle \left \langle \omega _{0_{-}}\right \vert
\right) $ \\ 
& $+$ & $\left \vert \omega _{\eta _{+}}\right \rangle \Gamma _{\alpha
}^{\eta _{+}\xi _{-}}\left \langle \omega _{\xi _{-}}\right \vert \left(
\left \vert \omega _{0_{+}}\right \rangle \left \langle \omega _{\beta
_{+}}\right \vert +\left \vert \omega _{\delta _{+}}\right \rangle \Gamma
_{\beta }^{\delta _{+}\gamma _{-}}\left \langle \omega _{\gamma _{-}}\right
\vert +\left \vert \omega _{\beta _{-}}\right \rangle \left \langle \omega
_{0_{-}}\right \vert \right) $ \\ 
& $+$ & $\left \vert \omega _{\alpha _{-}}\right \rangle \left \langle
\omega _{0_{-}}\right \vert \left( \left \vert \omega _{0_{+}}\right \rangle
\left \langle \omega _{\beta _{+}}\right \vert +\left \vert \omega _{\delta
_{+}}\right \rangle \Gamma _{\beta }^{\delta _{+}\gamma _{-}}\left \langle
\omega _{\gamma _{-}}\right \vert +\left \vert \omega _{\beta _{-}}\right
\rangle \left \langle \omega _{0_{-}}\right \vert \right) $%
\end{tabular}%
\end{equation}%
Moreover, using properties of the weight vector states, in particular the
orthogonality feature exhibited by (\ref{56}), we can bring the above
expression into the following simple form%
\begin{equation}
X_{\alpha }X_{\beta }=\left \vert \omega _{0_{+}}\right \rangle \left
\langle \omega _{\alpha _{+}}|\omega _{\delta _{+}}\right \rangle \Gamma
_{\beta }^{\delta _{+}\gamma _{-}}\left \langle \omega _{\gamma _{-}}\right
\vert +\left \vert \omega _{\eta _{+}}\right \rangle \Gamma _{\alpha }^{\eta
_{+}\xi _{-}}\left \langle \omega _{\xi _{-}}|\omega _{\beta _{-}}\right
\rangle \left \langle \omega _{0_{-}}\right \vert
\end{equation}%
To make the calculation covariant, it is interesting to use the metric of
the representation $\mathbf{27}_{+1/2}$ to set $\left \langle \omega
_{\alpha _{+}}|\omega _{\delta _{+}}\right \rangle \Gamma _{\beta }^{\delta
_{+}\gamma _{-}}=\Gamma _{\alpha _{+}\beta }^{\gamma _{-}}$ and also use the
metric of the representation $\mathbf{27}_{-1/2}$ to do the same thing for $%
\Gamma _{\alpha }^{\eta _{+}\xi _{-}}\left \langle \omega _{\xi _{-}}|\omega
_{\beta _{-}}\right \rangle =\Gamma _{\alpha \beta _{-}}^{\eta _{+}}.$ In
doing so, the above $X_{\alpha }X_{\beta }$ is further reduced as follows $%
X_{\alpha }X_{\beta }=\left \vert \omega _{0_{+}}\right \rangle \Gamma
_{\alpha _{+}\beta }^{\gamma _{-}}\left \langle \omega _{\gamma
_{-}}\right
\vert +\left \vert \omega _{\eta _{+}}\right \rangle \Gamma
_{\alpha \beta _{-}}^{\eta _{+}}\left \langle \omega _{0_{-}}\right \vert $.
Substituting the obtained expression of $X_{\alpha }X_{\beta }$ into the
expansion of $X^{2}$ namely $b^{\alpha }b^{\beta }X_{\alpha }X_{\beta }$, we
end up with the following result 
\begin{equation}
X^{2}=2S^{\gamma _{-}}\left \vert \omega _{0_{+}}\right \rangle \left
\langle \omega _{\gamma _{-}}\right \vert +2S^{\eta _{+}}\left \vert \omega
_{\eta _{+}}\right \rangle \left \langle \omega _{0_{-}}\right \vert
\end{equation}%
where for commodity we have set 
\begin{equation}
S^{\gamma _{-}}=\frac{1}{2}b^{\alpha }\Gamma _{\alpha _{+}\beta }^{\gamma
_{-}}b^{\beta }\qquad ,\qquad S^{\eta _{+}}=\frac{1}{2}b^{\alpha }\Gamma
_{\alpha \beta _{-}}^{\eta _{+}}b^{\beta }  \label{2s}
\end{equation}%
which are quadratic into the $b^{\alpha }$'s. The $S^{\gamma _{-}}$ couples
the $\omega _{0_{+}}$ with $\omega _{\gamma _{-}}$ whereas the $S^{\eta
_{+}} $ couples the $\omega _{\eta _{+}}$ and $\omega _{0_{-}}$.\textbf{\ }%
Here an interesting question emerges, it concerns the proof of the equality
of $S^{\gamma _{-}}$ and $S^{\eta _{+}}$ as suggested by physical and
representation theory arguments; see also Figure \textbf{6}. We will turn to
answering positively this question later on. Before that notice that similar
analysis can be done for the calculation of $Y^{2}=Y^{\alpha }Y^{\beta
}c_{\alpha }c_{\beta }$ with $Y^{\alpha }$ given by (\ref{YX}). We find%
\begin{equation}
Y^{2}=2R_{\alpha _{+}}\left \vert \omega _{0_{-}}\right \rangle \left
\langle \omega ^{\alpha _{+}}\right \vert +2R_{\eta _{-}}\left \vert \omega
^{\eta _{-}}\right \rangle \left \langle \omega _{0_{+}}\right \vert
\end{equation}%
where we have set%
\begin{equation}
R_{\alpha _{-}}=\frac{1}{2}c_{\gamma }\bar{\Gamma}_{\alpha _{-}}^{\gamma
\delta _{+}}c_{\delta }\qquad ,\qquad R_{\alpha _{+}}=\frac{1}{2}c_{\gamma }%
\bar{\Gamma}_{\alpha _{+}}^{\gamma _{-}\delta }c_{\delta }  \label{R}
\end{equation}%
These $R_{\alpha }$'s are quadratic in the c's. The $R_{\alpha _{+}}$
couples $\omega _{0_{-}}$ and the $\omega ^{\alpha _{+}}$ while the $%
R_{\alpha _{-}}$ couples $\omega ^{\eta _{-}}$ and $\omega _{0_{+}}.$ Here
also the same question asked for the two equalities in (\ref{2s}) can be
asked for (\ref{R}). Are the $R_{\alpha _{-}}$ and the $R_{\alpha _{+}}$
equal? The answer is affirmative; it is demonstrated below by considering
the associative property $X^{2}.X=X.X^{2}$.

$\bullet $ \emph{Calculation of }$X^{3}$ \emph{and} $Y^{3}:$ To perform the
calculation of $X^{3},$ we can decompose it either like $X.X^{2}$ or as $%
X^{2}.X$; thanks to associativity. By using $X.X^{2}$ and the above results
for X and X$^{2},$ we can determine the explicit expression of $X^{3}$.
First, we have%
\begin{equation}
\begin{tabular}{lll}
$X^{3}$ & $=$ & $b^{\beta }\left( \left \vert \omega _{0_{+}}\right \rangle
\left \langle \omega _{\beta _{+}}\right \vert \right) \left( 2S^{\alpha
_{-}}\left \vert \omega _{0_{+}}\right \rangle \left \langle \omega _{\alpha
_{-}}\right \vert +2S^{\eta _{+}}\left \vert \omega _{\eta _{+}}\right
\rangle \left \langle \omega _{0_{-}}\right \vert \right) +$ \\ 
& $\ $ & $b^{\beta }\left( \left \vert \omega _{\delta _{+}}\right \rangle
\Gamma _{\beta }^{\delta _{+}\gamma _{-}}\left \langle \omega _{\gamma
_{-}}\right \vert \right) \left( 2S^{\alpha _{-}}\left \vert \omega
_{0_{+}}\right \rangle \left \langle \omega _{\alpha _{-}}\right \vert
+2S^{\eta _{+}}\left \vert \omega _{\eta _{+}}\right \rangle \left \langle
\omega _{0_{-}}\right \vert \right) +$ \\ 
&  & $b^{\beta }\left( \left \vert \omega _{\beta _{-}}\right \rangle \left
\langle \omega _{0_{-}}\right \vert \right) \left( 2S^{\alpha _{-}}\left
\vert \omega _{0_{+}}\right \rangle \left \langle \omega _{\alpha
_{-}}\right \vert +2S^{\eta _{+}}\left \vert \omega _{\eta _{+}}\right
\rangle \left \langle \omega _{0_{-}}\right \vert \right) $%
\end{tabular}%
\end{equation}%
Then, using the same properties mentioned previously, we can bring the above
relation to a simple form as follows $X^{3}=2b^{\beta }g_{\beta _{+}\eta
_{+}}S^{\eta _{+}}\left \vert \omega _{0_{+}}\right \rangle \left \langle
\omega _{0_{-}}\right \vert $ where we have set $g_{\beta _{+}\eta
_{+}}=\left \langle \omega _{\beta _{+}}|\omega _{\eta _{+}}\right \rangle .$
By putting $b_{\eta _{+}}=b^{\beta }g_{\beta _{+}\eta _{+}}$ or equivalently 
$S_{\beta _{+}}=g_{\beta _{+}\eta _{+}}S^{\eta _{+}}$, we can reduce the
above relation down to 
\begin{equation}
X^{3}=6\mathcal{E}\left \vert \omega _{0_{+}}\right \rangle \left \langle
\omega _{0_{-}}\right \vert \qquad ,\qquad \mathcal{E}=\frac{1}{6}\Gamma
_{\alpha \beta \gamma }b^{\alpha }b^{\beta }b^{\alpha }  \label{3X}
\end{equation}%
where we have set $\mathcal{E}=\frac{1}{3}b_{\eta _{+}}S^{\eta _{+}}$ which
can be also presented like $\mathcal{E}=\frac{1}{3}b^{\eta _{+}}S_{\eta
_{+}}.$ To get more insight into the algebraic property of the scalar $%
\mathcal{E}$, we re-calculate $X^{3}$ by using the factorisation $X^{2}.X,$
and compare it with the result obtained using $X.X^{2}$. We find $\mathcal{E}%
=\frac{1}{3}b_{\eta _{-}}S^{\eta _{-}}$ or equivalently $\mathcal{E}=\frac{1%
}{3}b^{\eta _{-}}S_{\eta _{-}}.$ Comparing with the expression obtained
before namely $\mathcal{E}=\frac{1}{3}b_{\eta _{+}}S^{\eta _{+}},$ we end up
with the equality $S^{\eta _{+}}=S^{\eta _{-}}.$ In eq(\ref{3X}), the $%
\mathcal{E}$ couples $\omega _{0_{+}}$ and $\omega _{0_{-}}$ and is cubic
into the $b$'s. To exhibit manifestly this cubic dependence, we substitute $%
S^{\eta _{+}}$ by its value (\ref{2s}). We end up with the following
expression $\mathcal{E}=\frac{1}{6}\Gamma _{\alpha \beta \gamma }b^{\alpha
}b^{\beta }b^{\alpha }$ which can be interpreted in terms of the trace of
the tensor product $27\times 27\times 27$. Similar calculations for $Y^{3}$
lead to%
\begin{equation}
Y^{3}=6\mathcal{F}\left \vert \omega _{0_{-}}\right \rangle \left \langle
\omega _{0_{+}}\right \vert \qquad ,\qquad \mathcal{F}=\frac{1}{6}\bar{\Gamma%
}^{\alpha \beta \gamma }c_{\alpha }c_{\beta }c_{\gamma }  \label{3Y}
\end{equation}%
with metric $g_{\beta _{-}\eta _{-}}=\left \langle \omega _{\beta
_{-}}|\omega _{\eta _{-}}\right \rangle $ and $\mathcal{F}=\frac{1}{3}%
c^{\eta _{-}}R_{\eta _{-}}$ where $R_{\eta _{-}}$\ is given by (\ref{R}).
Here also, we have $R_{\alpha _{+}}=R_{\alpha _{-}}$. By substituting $%
R_{\eta _{-}}$ by its value, we end up with $\mathcal{F}=\frac{1}{6}\bar{%
\Gamma}^{\alpha \beta \gamma }c_{\alpha }c_{\beta }c_{\gamma }$
corresponding to the trace of the tensor product $\overline{27}\times 
\overline{27}\times \overline{27}$. From the expressions of (\ref{3X}) and (%
\ref{3Y}), we learn that $X^{4}=Y^{4}=0$ due to the decomposition (\ref{56}%
). So the expansion of the exponentials $e^{X}$ and $e^{Y}$ terminates at
the fourth order. This is a property of the E$_{7}$ theory. 
\begin{equation*}
\text{ \  \  \ }
\end{equation*}

\end{document}